\documentstyle[aps,epsf,epsfig,12pt]{revtex}
\draft

\begin{document}
\baselineskip 5mm

\title{Spin-Polaron approach to the Fermi surface evolution in the normal
state of cuprates}
\author{A.F.Barabanov$^*$, R.Hayn$^{+}$, A.A.Kovalev$^*$, O.V.Urazaev$^*$, 
and A.M.Belemouk$^*$  
\\
$^*${\normalsize Institute for High Pressure Physics, 
142190 Troitsk, Moscow region, Russia}\\
$^{+}${\normalsize Institute for Solid State and Materials Research (IFW),} \\
{\normalsize P.O. Box 270016, D-01171 Dresden, Germany}\\
}
\date{}
\maketitle

\begin{abstract}
\baselineskip 5mm
A semi-phenomenological approach to describe the evolution of Fermi surface
(FS) and electronic structure with doping is presented which is based on the
spin-fermion model. The doping is simulated by a frustration term in the spin 
Hamiltonian and the complex internal structure of the spin-polaron
quasiparticle is taken into account by a superposition of spin-polaron states
with different radii. By calculating the spectrum, the spectral weights and
the FS we find a rather drastic change of the electronic structure with doping
which explains many photoemission data, i.e.\ the isotropic band bottom and
the remnant FS of undoped cuprates, the large FS and the extended saddle point 
of optimally doped compounds and the pseudogap in the underdoped samples. \\
PACS number(s): 75.50.Ee, 74.20.Mn, 71.38.+i, 75.30.Mb
\end{abstract}

\section{\bf \ Introduction}
The evolution of Fermi surface (FS) and electronic spectrum in the normal
state of 
high temperature superconductors are intensively studied at present. 
Recent angle-resolved photoemission
spectroscopy (ARPES) studies indicate different dispersion relations in the
insulating and optimally doped cuprates. The undoped 
compounds \cite{wsmkk95,mdlpm96} show an isotropic band
bottom close to $N=(\pi/2,\pi/2)$ in momentum space, a large energy
difference between $N$ and $X=(\pi ,0)$, and a so-called remnant Fermi surface
\cite{rkfm98}
(that surface where the single particle spectral weight shows a sudden drop
of intensity). In the optimally doped compounds 
a flat band region, a large FS centered at $M=(\pi,\pi)$ 
and a so-called shadow Fermi surface due to antiferromagnetic (AFM)
spin correlations \cite{ks90} are found
\cite{togls92,gcdgl93,acg93,dskml93,ksdmp94,aos94,bgl00}. The flat band
region has the 
form of an extended saddle point which stretches in the direction 
$(\pi/2,0)$-$(\pi,0)$. The shadow FS resembles the main FS but is shifted by
the AFM vector ${\bf Q=(}\pi ,\pi {\bf ).}$ Furthermore, the underdoped
compounds demonstrate the existence of a high energy pseudogap near the
point $X$ with an energy of about $0.1\div 0.2$ eV  
\cite{mdlpm96,lsd96,dyc96,dnytr97}. Assuming a simple rigid band filling,
the isotropic minimum of the undoped compounds would suggest small hole
pockets around $N$. However, there are no clear experimental indications for
them \cite{dnytr97} and the Fermi surface seems to develop in an arc-like
fashion \cite{ndr98}. It is a real theoretical problem to reconcile that
seeming  
contradiction which can be attacked only by considering the spectral density
evolution as a whole. 

There are numerous studies of the hole spectrum in two dimensional (2D) doped
antiferromagnets using a wealth of theoretical models, like 
the generalized $t$-$J$, the effective three-band model, the Kondo-lattice or
the Hubbard model. Most of those studies are restricted to a very low hole
concentration using the exact diagonalization of small
clusters \cite{shsz90,djmbg90}, the 
quantum Monte Carlo technique \cite{dnh97}, the self-consistent Born
approximation (SCBA) \cite{klr89,mh91,tsm00} or a ``string'' ansatz for the
hole wave function \cite{eb90}.  One can start either
from the two-sublattice N\'eel-type state \cite{klr89,mh91,eb90,cdsu90} or
from the spin rotationally invariant spin liquid state \cite{bbzm96,hbs97}
and obtain qualitatively the same result: the hole motion takes place mainly
on one sublattice, i.e.\ the dispersion relation is dominated by an
effective hopping to the next nearest neighbors and the dispersion has its 
minimum at $N$ for small doping. There are much less attempts to explain the
Fermi surface evolution, like for instance \cite{cm97} which used a
diagrammatic method 
combined with a 
semi-phenomenological spin susceptibility.
The exact diagonalization studies (on 
clusters containing up to $16\div 20$ sites) at finite hole doping
\cite{eos97,kws98,ygl98} have difficulties to reconstruct
the FS and to investigate a sufficiently smooth variation of
doping due to the restricted set of ${\bf k}$-points.

The approach which shall be presented here is semi-phenomenological as
well. On one hand, we start with the spin-fermion model (the reduced form of
the three-band Emery model) that has a clear microscopic basis
\cite{e87,er88}. We also include direct oxygen-oxygen hopping terms which are 
rather important. On the other hand, we simulate the doping by a frustration
term $J_2$ in addition to the nearest neighbor exchange $J_1$ of the spin part 
in the 
Hamiltonian. The similarity between doping and frustration had been first
proposed in \cite{idg88} based on a similar decrease of the magnetic
correlation length. It is interesting to note that cluster calculations
indicated a considerable value of the frustration parameter $J_2/J_1\sim 0.1$
even for the undoped La$_2$CuO$_4$ \cite{amms89}. Of course there is no full
equivalence between doping and frustration. For 
example, the doped $t$-$J$ model and the frustrated $J_1$-$J_2$ model give
different results for the dynamical spin-spin structure factor and for the
spectrum of Raman scattering \cite{bgn91}. But numerical
calculations on finite lattices 
indicate the equivalence of the mentioned models if we are interested in the
static spin-spin correlation functions \cite{mdjr90}. And those quantities are 
especially relevant in our present Green's function method where the
spectrum $\varepsilon ({\bf k})$ is determined by the static spin-spin
correlation functions of the spin subsystem.

Previous studies of the frustrated
generalized $t$-$J$ \cite{hbs97} and the frustrated effective three-band
model \cite{bbzm96} demonstrated already the strong influence of frustration on
the spin-polaron spectrum. However, these investigations were performed only
within the local polaron approach (LPA) which cannot describe the influence of 
long range spin order on the excitation spectrum. Furthermore, only the bare
spectral weight was considered which could only give a hypothetical FS. 
Recently 
\cite{stm99} the frustrated $t-t^{\prime }-t^{\prime \prime }-J_1-J_2$ model
was discussed in the framework of the SCBA where the holes are treated as
spinless 
fermions and the spins as normal bosons \cite{mh91}. The resulting SCBA in
\cite 
{stm99} corresponds to a two sublattice approach in which the spectrum is
dictated by the 
symmetry of the reduced magnetic Brillouin zone (BZ) (e.g. the
points $\Gamma $ and $M$ are equivalent). However, such a symmetry is not 
reproduced in the ARPES data. On the contrary, the spherically symmetric
approach (in the spin space) is used in the present investigation to describe
the spin subsystem.  
As a result, the spectrum is periodic relative to the full
Brillouin zone (BZ) of the CuO$_2$ plane.

The distinctive feature of our present investigation consists in treating the
spin-polaron as a complex quasiparticle, i.e.\ as a 
superposition of spin polaron states with
different radii \cite{bkub00}. Our basis operators enclose the local 
polaron (bare holes and the Zhang-Rice polaron) and also its coupling to 
spin waves with quasimomentum ${\bf q}$ close to the antiferromagnetic wave
vector ${\bf Q}$. By introducing the complex structure of the spin-polaron one
may describe  the important
splitting of the lowest local polaron band. This allows, for example, to
reproduce for heavily underdoped compounds a sudden drop in the intensity
of ARPES peaks as ${\bf k}$ goes from $N$ to $M$. And we will show that it
reproduces also many of the other features of electronic structure evolution
mentioned above. 

\section{\bf \ The Hamiltonian and the method}

The main features of the hole motion in the ${\rm CuO}_2$ plane are described
by the model \cite{e87,er88}:

\begin{equation}  \label{r6-ham}
\hat H=\hat \tau +\hat J+\hat h\;,
\end{equation}
\begin{equation}  \label{r6-J}
\hat \tau =4\tau \sum_{{\bf R}}p_{{\bf R}}^{\dagger }\left( \frac 12+\tilde
S_{{\bf R}}\right) p_{{\bf R}}\;,
\end{equation}
\begin{equation}  \label{r3a-1}
\qquad \hat J=\frac{J_1}2\sum_{{\bf R},{\bf g}}{\bf S}_{{\bf R}}{\bf S}_{%
{\bf R}+{\bf g}}+\frac{J_2}2\sum_{{\bf R},{\bf d}}{\bf S}_{{\bf R}}{\bf S}_{%
{\bf R}+{\bf d}}
\end{equation}
where $J_1,J_2\!\!>0$ are the antiferromagnetic interactions between the
first-nearest neighbor (${\bf g=\pm g}_x,{\bf \pm g}_y$) and the
second-nearest neighbor (${\bf d=\pm g}_x{\bf \pm g}_y$) on a square
lattice. The direct oxygen-oxygen hopping is included by 
\begin{equation}  \label{r6-hh}
\hat h=-h\sum_{{\bf R}}\left[ c_{{\bf R+a}_x}^{\dagger }\left( c_{{\bf R+a}%
_y}+c_{{\bf R-a}_y}+c_{{\bf R+g}_x{\bf +a}_y}+c_{{\bf R+g}_x-{\bf a}%
_y}\right) +h.c.\right] \;,
\end{equation}
with 
\begin{equation}  \label{r6-add1}
p_{{\bf R}}=\frac 12\sum_{{\bf a}}c_{{\bf R}+{\bf a}},\quad \tilde S_{{\bf R}%
}\equiv S_{{\bf R}}^\alpha \hat \sigma ^\alpha ,\quad \left\{ p_{{\bf R}},p_{%
{\bf R}^{\prime }}^{\dagger }\right\} =\delta _{{\bf R,R}^{\prime }}+\frac
14\sum_{{\bf g}}\delta _{{\bf R,R}^{\prime }+{\bf g}}\;,
\end{equation}
where ${\bf a}_{x,y}=\frac 12{\bf g}_{x,y}.$

Here and below the summation over repeated indexes is understood everywhere; 
$\{\ldots ,\ldots \},\ [\ldots ,\ldots ]$ stand for anticommutator and
commutator respectively; ${\bf g}_{x,y}$ are basis vectors of a copper
square lattice ( $|{\bf g}|\equiv 1$), ${\bf R}+{\bf a}$ are four vectors of
O sites nearest to the Cu site ${\bf R}$; the operator $c_{{\bf R+a}%
}^{\dagger }$ creates a hole predominantly at the O site (the spin index is
dropped in order to simplify the notations); $\hat \sigma ^\alpha $ are the
Pauli matrices; the operator ${\bf S}$ represents the localized spin on the
copper site. We do not introduce the explicit relative phases of $p$- and $d$
-orbitals since they can be transformed out by redefining the operators with
phase factors $\exp \left( \imath {\bf QR}\right) $. In order to compare our
results with other authors and with experiment we restore these phases by
changing in the final results ${\bf k}\rightarrow {\bf k}^{\prime }={\bf k}-%
{\bf Q}$. The parameter $\tau $ is the hopping amplitude of oxygen holes
that takes into account the coupling of the hole motion with the copper spin
subsystem. For convenience, the exchange interactions  $J_1$ and $J_2$ can be
expressed in terms of the frustration parameter $p$:  
\begin{equation}  \label{r3a-2}
J_1=(1-p)\,J,\quad J_2=pJ,\quad 0\leq p\leq 1,\quad J>0 \; . 
\end{equation}
The frustration parameter $p$ has to be connected with the number $x$ of
holes per copper atom. We are guided by 
an estimate based on the single-band Hubbard model
with realistic values of $U/t\sim 5$ for $x=0.1$ where one finds a value of
$p\sim 0.1$.  Let us 
note that in the case of La$_{2-x}$Sr$_x$CuO$_4$ the spin
system of the CuO$_2$ plane looses its long-range order for $x>0.02$.

The adopted effective spin-fermion Hamiltonian can be obtained by canonical
transformation from the Emery model \cite{zo88} in the regime $U_d\gg t\gg
\varepsilon $, $\varepsilon =\varepsilon _p-\varepsilon _d>0$, where $U_d$ is
the Coulomb repulsion on copper sites. The values of the energy
parameter set $\varepsilon $, $t$, U$_d$ may be
extracted from band structure or cluster calculations. Two typical
sets are: $\varepsilon =\varepsilon_p-\varepsilon_d= 3.6$ eV, $t=1.3$ eV,  
$U_d=10.5$ eV, \cite{hsc89}, or $\varepsilon =2.0$ eV, $t=1.0$ eV $U_d=8.0$
eV \cite{ptoyi88,mms88}. Taking into account that $\tau \sim 
t^2/\varepsilon $, $J\sim 4\tau ({{t/\varepsilon )}^2\ }$we adopt the
parameter values $\tau \sim 0.4$ eV, $J=0.4\tau$ and $h=0.4\tau $. Below we
take $\tau$ as the unit of energy. We consider a spin-liquid state with 
spin rotational symmetry for the copper subsystem, e.g. the spin-spin
correlation functions satisfy the relation $C_{{\bf r}}\equiv \left\langle
S_{{\bf R}}^\alpha S_{{\bf R}+{\bf r}}^\alpha \right\rangle =3\left\langle
S_{{\bf R}}^{x\left( y,z\right) }S_{{\bf R}+{\bf r}}^{x\left( y,z\right)
}\right\rangle ,\quad <S_{{\bf R}}^\alpha >=0.$

To treat the problem of a spin-polaron in the projection method approach we
introduce a finite set of basis operators connected with each cell ${\bf R}$%
: $A_{{\bf R},i}^{+}$ (with spin $\sigma $), where $i$ is the number of the
operator, $i\leq N$. The set contains the creation of a hole (with spin $%
\sigma $) and other operators which describe a hole connected with spin
excitations at different distances. The choice of the set is dictated by the
physical sense of the problem and will be explained in detail below. To
calculate the spectral function of bare holes we take the sum of two
contributions $c_{{\bf R+a}_x}^{+}$ and $c_{{\bf R+a}_y}^{+}$ that are
related with two possible positions of an oxygen hole in the unit cell.

As usual we introduce the retarded two-time Green's functions $G_{ij}(t,{\bf %
k})$ defined in terms of the Fourier transforms $A_{{\bf k},i}$ of the
operators $A_{{\bf R},i}$:
\begin{equation}  \label{r4b-gr}
G_{ij}(t,{\bf k})\equiv \left\langle A_{{\bf k},i}^{}(t)\big\vert A_{{\bf k}%
,j}^{+}(0)\right\rangle =-i\Theta (t)\left\langle \left\{ A_{{\bf k}%
,i}^{}(t)\;,A_{{\bf k},j}^{+}(0)\right\} \right\rangle \;,
\end{equation}
\[
A_{{\bf k},j}={\frac 1{{\sqrt{N}}}}\sum_{{\bf R}}{\rm e}^{i{\bf kR}}A_{{\bf R%
},j}\;\qquad i,j=1\div N\;. 
\]
The equations of motion for the Fourier transforms of the Green's functions
have the form
\begin{equation}  \label{r4b-green}
\begin{array}{l}
\omega {\left\langle A_{{\bf k},i}^{}\big\vert A_{{\bf k},j}^{+}\right%
\rangle }_\omega =K_{ij}+{\left\langle B_{{\bf k},i}^{}\big\vert A_{{\bf k}%
,j}^{+}\right\rangle }_\omega \\ 
K_{ij}({\bf k})=\left\langle \left\{ A_{{\bf k},i}^{}\;,A_{{\bf k}%
,j}^{+}\right\} \right\rangle \;,\qquad B_{{\bf k},i}^{}=\left[ A_{{\bf k}%
,i}^{}\;,H\right] \; . 
\end{array}
\end{equation}
In the projection technique we approximate the new operators $B_{{\bf k},i}$
by their projections on the space $\left\{ A_{{\bf k},i}\right\} $ of the
basis operators: 
\begin{equation}  \label{r4b-matrix}
B_{{\bf k},i}\simeq \sum_l{L}_{il}({\bf k})A_{{\bf k},l}\;,\quad {L}({\bf k}%
)=D({\bf k})K^{-1}\;,\quad {D}_{ij}({\bf k})=\left\langle \left\{ B_{{\bf k}%
,i}^{}\;,A_{{\bf k},j}^{+}\right\} \right\rangle .
\end{equation}
After substitution of the approximate expressions for the operators $B_{{\bf
k}%
,i}$ in (\ref{r4b-matrix}) into the equations of motion (\ref{r4b-green}),
the equation system (\ref{r4b-green}) for Green's functions ${\left\langle
A_{{\bf k},i}^{}\big\vert
A_{{\bf k},j}^{+}\right\rangle }_\omega $ becomes closed and can be
presented in the matrix form 
\begin{equation}  \label{m1}
\left( \omega E-DK^{-1}\right) G=K \; , 
\end{equation}
where $E$ is the unit matrix.
The quasiparticle spectrum $\varepsilon ({\bf k})$ is determined by the
poles of the Green's function $G$ and can be derived from the condition 
\[
det|K\varepsilon ({\bf k})-D|=0 \; . 
\]

\section{\bf The basis operators and approximations}

At first we choose a set of three basis operators which describe local
spin-polaron states 
\begin{equation}  \label{p1}
A_{{\bf R},1}=\frac 12(c_{{\bf R+a}_x}+c_{{\bf R-a}_x}) \; ;
\quad A_{{\bf R},2}=\frac 12(c_{{\bf R+a}_y}+c_{{\bf R-a}_y}) \; ; 
\quad A_{{\bf R},3}=\tilde S_{{\bf R}}p_{{\bf R}} \; . 
\end{equation}
The combinations of them constitute the Zhang-Rice polaron and bare hole
states, 
in particular:
\begin{equation}  \label{p2}
{c_{{\bf k},x}^{}}=\frac 2{1+\exp (i k_x)}A_{{\bf k},1} \; ; 
\qquad {c_{{\bf k},y}^{}}=\frac 2{1+\exp (i k_y)}A_{{\bf k},2} \; ; 
\qquad {c_{{\bf k},x(y)}^{}=
\frac1{{\sqrt{N}}}}\sum_{{\bf R}}{\rm e}^{i{\bf kR}}{c_{{\bf R+a}_{x(y)}}^{}}
\; . 
\end{equation}
The spectrum of elementary excitations in the framework of the local polaron
approach (LPA) was investigated previously \cite{bbzm96}. In particular it
was pointed out that the frustration in spin subsystem and oxygen-oxygen
hopping can explain the appearance of an extended saddle-point in the
spectrum. 
The main shortcoming of LPA consists in the fact that the spectrum of
elementary excitations depends only on the short range spin-spin correlation
functions even at zero temperature. This means that LPA does not describe the 
influence of spin long range order on the excitation spectrum.

Treating the periodic Kondo problem on a square lattice in the two sublattice
spin structure, Schrieffer \cite{s95} emphasized the crucial importance of
taking into account coherence factors related to the existence of long
range order. In this model the band electrons (holes) are coupled to localized
spins and the simplest Hamiltonian has the form
\begin{equation}  \label{p3}
H_K=\sum_{{\bf R,g}}t_{{\bf g}}c_{{\bf R}+{\bf g}}^{+}c_{{\bf R}}+I\sum_{%
{\bf R}}c_{{\bf R}}^{+}\tilde S_{{\bf R}}c_{{\bf R}}+\frac 12J\sum_{{\bf R,g}%
}S_{{\bf R}+{\bf g}}^\alpha S_{{\bf R}}^\alpha
\end{equation}
where the insite exchange with constant $I$ is analogous to the $\hat \tau $
Hamiltonian (\ref{r6-J}) in our model.
In Ref.\ \cite{s95} the mean-field approach was taken for the N\'eel state at
$T=0$ and the spins were treated as classical vectors:
\begin{equation}  \label{p4}
S_{{\bf R}}^\alpha =\delta _{\alpha z}S_0e^{i{\bf QR}},\quad S_0=const.\quad
\end{equation}
In this approximation the Kondo-interaction Hamiltonian (\ref{p3}) takes the
form of a potential energy with doubled period. As a result, this
interaction leads to the hybridization of bare particle states with momenta 
${\bf k}$ and ${\bf k}+{\bf Q}$ and one should perform a standard $u-v$
transformation to take into account this hybridization from the very
beginning. In the adopted N\'eel state the amplitude ${S}_{{\bf Q}}$ of the
spin wave with ${\bf q}={\bf Q}${\bf \ }(the ${\bf Q}$-wave) has a 
macroscopic large value and has properties analogous to the amplitude of
a Bose particle with zero momentum in the superfluid Bose-gas. As a result, 
this amplitude can be treated as a $c-$number for many problems. Then the
hybridization corresponds to coupling of the ${\bf Q}$-wave to local
electron states. But it does not represent new states and leads only to
mixing of states with momenta ${\bf k}$ and ${\bf k}+{\bf Q}$.

The distinctive feature of the present investigation consists in considering
the one-particle motion on the background of the spherically symmetric spin
subsystem state. In this background the average value $<{\bf S}_{{\bf Q}}>=0$
and the above mentioned treatment fails. At $T=0$ and without frustration
the only value which can be treated as a macroscopic one is $<{\bf S}_{{\bf
Q}}%
{\bf S}_{{\bf Q}}>$. Then the coupling of a local state to ${\bf S}_{{\bf Q}%
} $ corresponds to a new delocalized spin polaron states -- local polaron
states (\ref{p1}) dressed by antiferromagnetic spin wave ${\bf S}_{{\bf
Q}}$. For the model under discussion at $T=0$ such states were introduced in  
\cite{bmzu97} and have the form 
\begin{equation}  \label{p5}
\tilde Q_{{\bf r}}A_{{\bf R},i} \; ; \quad i=1,2,3
\end{equation}
\begin{equation}  \label{p6}
\tilde Q_{{\bf R}}=e^{-i{\bf QR}}\tilde S_{{\bf Q}} \; ; 
\quad \tilde S_{{\bf Q}}=S_{{\bf Q}}^\alpha \hat \sigma ^\alpha \; ; 
\quad {\bf S}_{{\bf Q}}=N^{-1}\sum_{{\bf R}_1}
e^{i{\bf QR}_1}{\bf S}_{{\bf R}_1} \; . 
\end{equation}
It was shown in \cite{bmzu97} that it is important to take into account the 
quantum nature of the spin ${\bf Q}$-wave because the transitions between
the local polaron states and the delocalized polaron states lead to a 
splitting of the lowest LPA bands and change their properties in an essential
way. Only the complex spin-polaron approach, i.e. that approach containing the 
local spin-polaron operators {\em plus} those operators dressed by the ${\bf
Q}$-wave, can be compared with a Green's function calculation for the bare
hole spectral function $A_h({\bf k},\omega)$ taking into account the
spin-polaron damping by means of the SCBA \cite{khbm98}. One can see that the
SCBA quasiparticle peak and its intensity are well given by the lowest band of 
the complex spin-polaron approach, whereas the excited bands reflect the
non-coherent part of $A_h({\bf k},\omega)$. 

In the present paper we treat the model at nonzero temperatures $T$ and
frustration $p$. In this case, the
spin subsystem looses its long-range order and the average value $<{\bf S}_{%
{\bf Q}}{\bf S}_{{\bf Q}}>$ is zero. So one must introduce new operators in
order to preserve the complex spin-polaron results and to take into account
a finite spin 
correlation length $\xi $. It is natural to generalize the above choice of
operators 
by introducing local polaron states 
dressed by spin waves with momenta ${\bf q}$ close to the
antiferromagnetic vector ${\bf Q}$ (for such ${\bf q}$ the spin- spin
structure factor is sharply peaked even at $T,\ p\neq 0$). These operators
can be taken in the form: 
\begin{equation}  \label{p7}
A_{{\bf R},4}=\tilde Q_{{\bf R}}^{(\Omega )}A_{{\bf R},1};\quad A_{{\bf R}%
,5}=\tilde Q_{{\bf R}}^{(\Omega )}A_{{\bf R},2};\quad A_{{\bf R},6}=\tilde
Q_{{\bf R}}^{(\Omega )}A_{{\bf R},3}
\end{equation}
\begin{equation}  \label{p8}
\tilde Q_{{\bf R}}^{(\Omega )}=N^{-1}\sum_{{\bf \rho ,q\in \Omega }}e^{i{\bf %
q\rho }}\tilde S_{{\bf R+\rho }};\quad {\bf \Omega }=\{{\bf q},|\pm \pi
-q_{x,y}|<\kappa _0\}
\end{equation}
Here ${\bf \Omega }$ is the square region around ${\bf Q}$ and around
equivalent 
points (described by the four squares $\Omega =\kappa
_0\times \kappa _0$ related to the corners of the first Brillouin zone (BZ),
see Fig. 1.1a). The choice of the parameter $\kappa _0$ is dictated by the 
spin correlation length $\xi$ which depends on the frustration. 
In order to clarify our 
future choice of the parameter $\Omega(\kappa _0)$ for different
frustrations let us represent each of the operators in (\ref{p7}) (taking 
$A_{{\bf R},4}$ as an example) in the form
\begin{equation}  \label{p9}
A_{{\bf R},4}=\sum_{{\bf R}^{\prime}} \alpha ({\bf R}-{\bf R}^{\prime})e^{i%
{\bf Q}({\bf R}-{\bf R}^{\prime})} \,\tilde S_{{\bf R}^{\prime}}A_{{\bf R},1}
\end{equation}
where
\begin{equation}
\alpha ({\bf l})=\int_{-\kappa _0}^{\kappa _0}\int_{-\kappa _0}^{\kappa _0}%
\frac{d^2k}{(2\pi )^2}e^{-i{\bf kl}}
\end{equation}
The absolute value of $\alpha ({\bf R}-{\bf R}^{\prime})$ depends only on
the absolute value of the difference ${\bf l}={\bf R}-{\bf R}^{\prime}$ and
decreases with the increase of $l$. This dependence $\tilde \alpha ({\bf l}%
)=\left| \alpha ({\bf l})/\alpha ({\bf 0})\right| $ has the form 
\begin{equation}  \label{p10}
\tilde \alpha ({\bf l})=\left| \sin (l_x\kappa _0)\sin (l_y\kappa
_0)/(l_xl_y\Omega )\right|
\end{equation}
The value $l_0$ , which satisfies the condition
\begin{equation}  \label{p11}
l_0\kappa _0=\pi ,
\end{equation}
can be qualitatively treated as a coupling radius of the local polaron which
defines the range of coupling to the long distance spin correlations, since $%
\tilde \alpha ({\bf l})\sim 1$ for $l<l_0$ and $\tilde \alpha ({\bf l})\ll 1$
for $l\gg l_0$. On the other hand, the coupling radius of a spin polaron
must be of the order of the spin correlation length $l_0\sim \xi $. This
qualitative estimation leads to the choice $\Omega =\pi
^2/\xi ^2,\ \kappa _0=\pi /\xi $. Such a choice of $\Omega $ gives a correct
description of two limits. If $\xi $ tends to infinity (the spin system has
long range order, $T=0$), the coupling radius of polaron also tends to
infinity, $\kappa _0$ tends to zero and $\tilde Q_{{\bf R}}^{(\Omega )}$ (%
\ref{p8}) is transformed to $\tilde Q_{{\bf R}}$ from (\ref{p6}).  It is clear
that in the opposite limit of high 
temperature, when the spin correlation between neighboring sites is small,
the system must be described by local spin polarons. As it is seen from (\ref
{p11}) if $\xi $ $(\sim l_0)$ tends to unity the value $\kappa _0$ is close
to $\pi $. In this case $\alpha ({\bf R}-{\bf R}^{\prime })=\delta _{{\bf R,R%
}^{\prime }}$ and the operators $A_{{\bf R},i},i=4,5,6$ represent local spin
polarons close to polarons $A_{{\bf R},i},i=1,2,3$ (in particular, for 
$\kappa _0=\pi$,  $A_{{\bf R},6}$ is the linear combination of $A_{{\bf
R},1}$, $A_{{\bf R},2}$, and $A_{{\bf R},3}$). As a
result, we have a continuous description of the model from a finite
temperature to zero one.

The spin correlation length was determined taking the spin wave
spectrum which was calculated for the frustrated Heisenberg model in the
framework of the spherically symmetric approach \cite{bb94a}. If the spin
system is not 
far from the N\'eel phase, $\xi $ may be determined using the expansion of
the spectrum close to the AFM point ${\bf Q}$ \cite{shim93}, which will be
denoted by 
$\xi _{{\bf Q}}$. Then we use the criterion $\kappa _0=\pi /\xi _{{\bf Q}%
}.$ The explicit expressions of spectrum and $\xi _{{\bf Q}}$ are given in
Appendix A.
The choice of the parameter $\kappa _0$ is different when 
$\xi _{{\bf Q}}$ becomes small. The value of $\xi _{{\bf Q}}$ 
depends strongly on the frustration parameter $p$. For $p\geq 0.15$ we have
$\xi _{{\bf Q}%
}\leq 2\div 3$ ($\kappa _0=\pi /\xi _{{\bf Q}}\leq 0.35\pi $) and for such a 
frustration the real value of $\xi $ can essentially differ from $\xi _{{\bf
Q}}$. 
Indeed, for the frustrated case the spin-spin correlation functions $C_{{\bf %
R}}=<{\bf S}_{{\bf R}_0}{\bf S}_{{\bf R}_0+{\bf R}}>$ have the following
dependence on ${\bf R=}n_x{\bf g}_x+n_y{\bf g}_y$: 
\begin{equation}
C_{{\bf R}}=m_1({\bf R})(-1)^{n_x+n_y}+m_2({\bf R})[(-1)^{n_x}+(-1)^{n_y}]
\label{p12}
\end{equation}
where $m_1({\bf R})\gg m_2({\bf R})$ for $p\ll 1$ (N\'eel type phase) and
$m_1(%
{\bf R})\ll m_2({\bf R})$ for $p$ close to unity (stripe phase with gapless
spectrum at points ${\bf Q}_1=(\pm \pi ,0),(0,\pm \pi )$). The values $m_1(%
{\bf R})$ and $m_2({\bf R})$ are dictated by the spin spectrum gaps in
the points ${\bf Q}$ and ${\bf Q}_1.$ For intermediate values of frustration $%
0.15\leq p\leq 0.55$ these gaps of the spin spectrum are comparable, and in
this $p$-interval it is meaningless to determine $\xi $ by the expansion of
the spectrum near the points ${\bf Q}$ or ${\bf Q}_1$. On the other hand, a
local 
polaron structure is taken into account by the operators $A_{{\bf R},1}$, 
$A_{{\bf R},2}$, $A_{{\bf R},3}$ and the operators $A_{{\bf R},4}$, 
$A_{{\bf R},5}$, $A_{{\bf R},6}$ 
are introduced as candidates to describe polaron states with
large or intermediate radius $l_0=2\div 3$. This means that we should fix $%
\kappa _0\simeq 0.35\pi $ for $p\geq 0.15$ ($\kappa _0=\pi /l_0,\ l_0=2\div
3)\ $and take $\kappa _0=\pi /\xi _{{\bf Q}}$ for $p<0.15$. The calculated
values of $\xi _{{\bf Q}}$ and the adopted below relation between the values
of the parameter $\kappa _0$ and different frustrations are represented in
Table~1.

To find the spectrum of excitations in the framework of the operators
(\ref{p1}),  
(\ref{p7}) we calculate the matrix elements for matrices $D$ and $K$.
The calculation of elements of the arrays $D$ and $K$ is usually lengthy and
it involves the calculation of complex commutators for the operators $A_{%
{\bf k},i}$ and $B_{{\bf k},i}$. Some of these commutators cannot be
expressed explicitly by two-site Green's functions and some approximations
are needed. The matrix elements are considerably simplified for a one-hole
approach which can be adopted due to treating the problem in the limit of low
doping (we will consider doping $n<0.2$). Then they are expressed in terms
of two- and multi-site correlation functions of spin operators (in the cases
considered below --- two-, three-, four- and five-site correlation
functions). By taking into account the spherical symmetry the three-site
correlators can be reduced to two-site correlators, and five-site
correlators can be reduced to four-site correlators. The four-site
correlators can be reduced to the form
\[
V_{{\bf R}_1{\bf R}_2{\bf R}_3{\bf R}_4}=\left\langle \left( {\bf S}_{{\bf R}%
_1}{\bf S}_{{\bf R}_2}\right) \,\left( {\bf S}_{{\bf R}_3}{\bf S}_{{\bf R}%
4}\right) \right\rangle , 
\]
where all sites are supposed to be different. To calculate such correlators
an approximation \cite{t88} is used:
\[
V_{{\bf R}_1{\bf R}_2{\bf R}_3{\bf R}_4}=C_{R_{12}}C_{R_{34}}+\frac
13C_{R_{13}}C_{R_{24}}+\frac 13C_{R_{14}}C_{R_{23}}. 
\]
As a result the matrix elements are expressed over the static spin-spin
structure factor $C_{{\bf q}}$ (Fourier transform of $C_{{\bf R}})$. The
explicit form of $K$ and $D$ matrix elements are reproduced in Appendix B. 
Typical sums in the matrix elements have the form:
\begin{equation}
u_g=N^{-1}\sum_{{\bf q\in \Omega }}e^{i{\bf qg}}C_{{\bf q}},\qquad
W_{g1}^{(J)}=N^{-2}\sum\limits_{{\bf q}_1{\bf ,q}_2{\bf \in \Omega }}e^{i%
{\bf q}_2{\bf g}}C_{{\bf q}_{_1}-{\bf q}_{_2}}C_{{\bf q}_{_2}}  \label{p13}
\; . 
\end{equation}
The expressions analogous to $u_g$ and $W_{g1}^{(J)}$ contain one and two
summations on ${\bf q\in \Omega }$. Each sum on ${\bf q}$ is proportional
to $\Omega /\pi ^2$ which is a small parameter in our approximation: as
mentioned above $\Omega /\pi ^2=\kappa _0^2/\pi ^2\leq 0.1$ (see Table 1).
This smallness allows to justify the approximation which consists in
neglecting some terms proportional to $\Omega ^2$.

\section{Results and discussion}

After solving the system (\ref{m1}) for the chosen set of operators $A_{%
{\bf R},i}$  the resulting Green's functions have the form 
\begin{equation}
\label{r1}G_{i,j}(\omega ,{\bf k})=\sum_{l=1}^6\frac{z_{(i,j)}^{(l)}({\bf k})%
}{\omega -\varepsilon _l({\bf k})},\qquad i,j=1\div 6 \; . 
\end{equation}
According to (\ref{p2}) the residues $z_{(1,1)}^{(l)}({\bf k})$,
$z_{(2,2)}^{(l)}({\bf k})$ determine the spectral weights 
$n_{h,\sigma}^{(l)}({\bf k})$ of bare oxygen holes with fixed
spin $\sigma $ and momentum ${\bf k}$ in the quasiparticle state $|{\bf k%
},\sigma ,l\rangle $ of the quasiparticle band $\varepsilon _l({\bf k})$: 
\begin{equation}
\label{r2}n_{h,\sigma }^{(l)}({\bf k})=\left( \frac 2{1+\cos
(k_x)}\right) z_{(1,1)}^{(l)}({\bf k})+\left( \frac 2{1+\cos (k_y)}\right) 
z_{(2,2)}^{(l)}({\bf k}) \; . 
\end{equation}
Let us remind that the bare hole spectral weight satisfies the sum rule $%
\sum_ln_{h,\sigma }^{(l)}({\bf k})=2$ and the maximum number of holes per
cell is equal to four despite the presence of six bands. This means that in
this model the Luttinger theorem is not fulfilled.

We present the results of the spin-polaron spectrum calculation for a
temperature $T=0.2J$. Let us mention that the spectrum has a weak temperature
dependence 
up to a temperature $T\sim 0.4J$. Since we are interested in relatively small
doping values $x\leq 0.2$ we represent results for the two lowest bands 
$l=1,2$ only. In the high doping regime our approach may be insufficient 
since it is 
based on the one hole approximation and does not take into account the 
interaction between polarons.

In Fig.\ 1 the spectrum $\varepsilon _1({\bf k})$ and the bare hole spectral
weight $n_{h,\sigma }^{(1)}({\bf k})$ of the lowest band for 
different values of the frustration parameter, $p=0.05$, 0.1, 0.13, 0.15, 0.2
and 0.25, are presented by
contour plots taking $\tau$ as the unit of energy.
As mentioned already, 
we suppose some qualitative equivalence between doping 
$x$ and frustration $p$. \\ 
\begin{minipage}{16cm}
\begin{minipage}[t]{5.3cm}
\begin{figure}
\mbox{\psfig{file=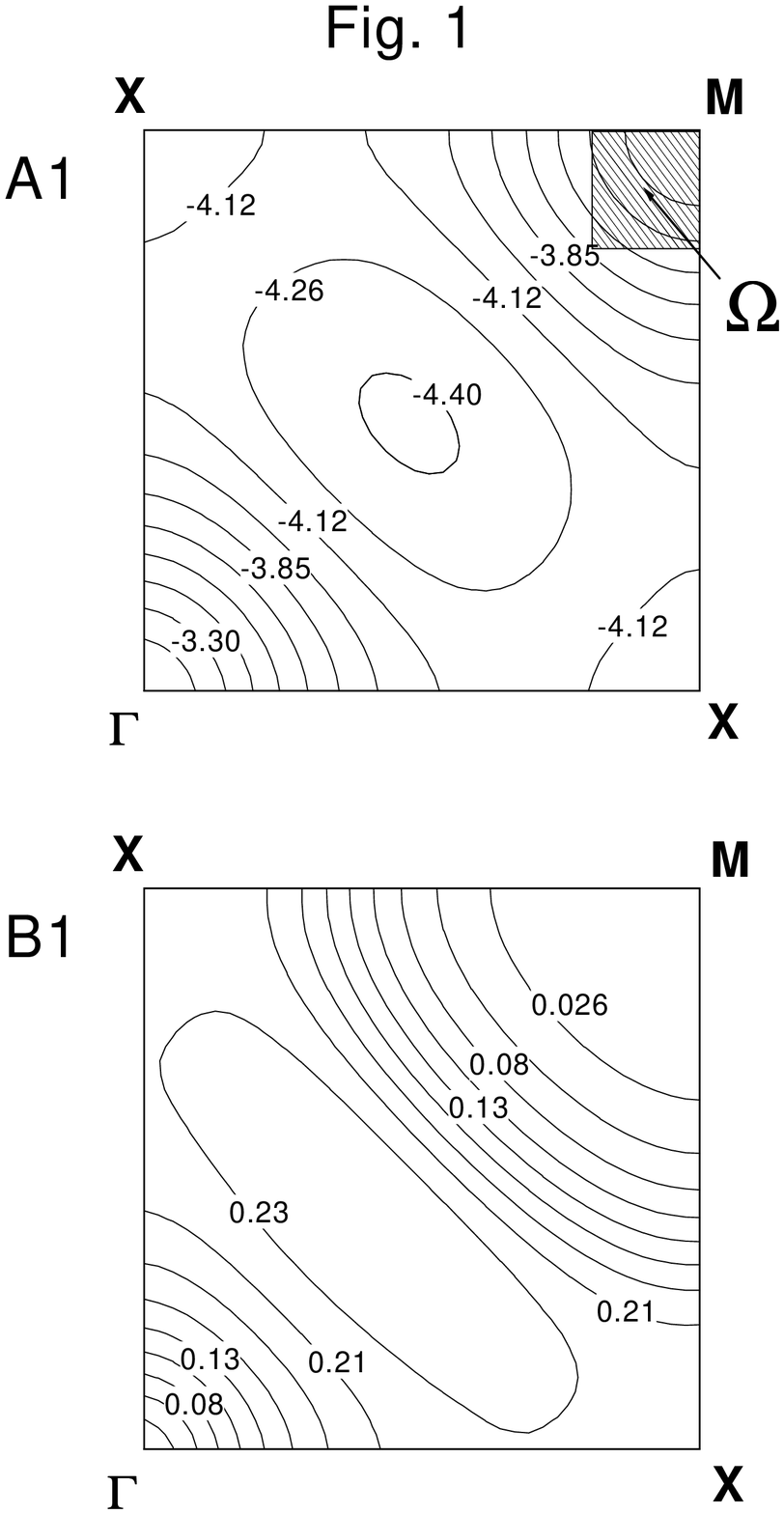,angle=0,width=5.3cm}}
\end{figure}
\end{minipage} 
\hspace*{-1cm}
\begin{minipage}[t]{5.3cm}
\begin{figure}
\mbox{\psfig{file=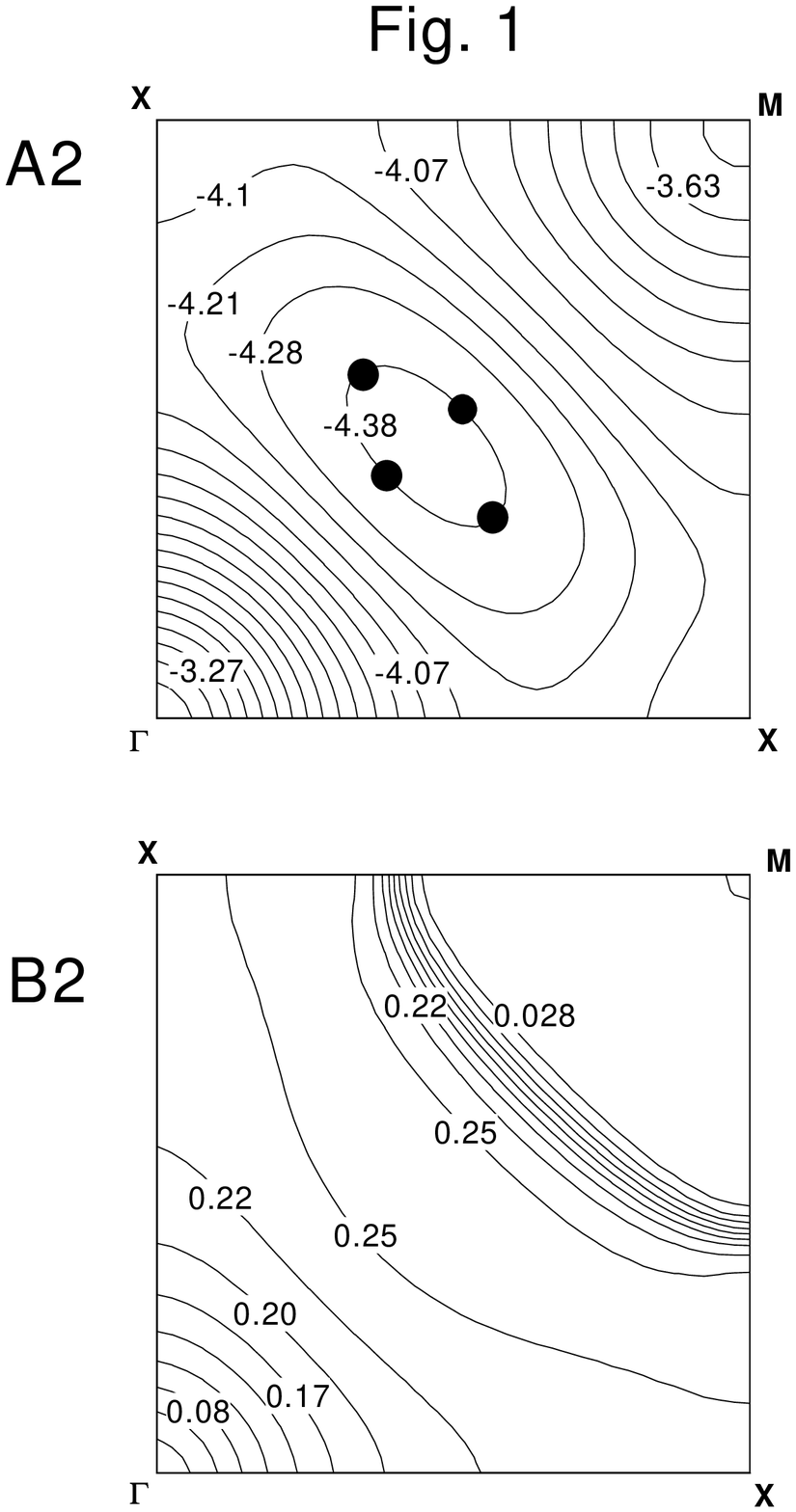,angle=0,width=5.3cm}}
\end{figure}
\end{minipage}
\hspace*{-1cm}
\begin{minipage}[t]{5.3cm}
\begin{figure}
\mbox{\psfig{file=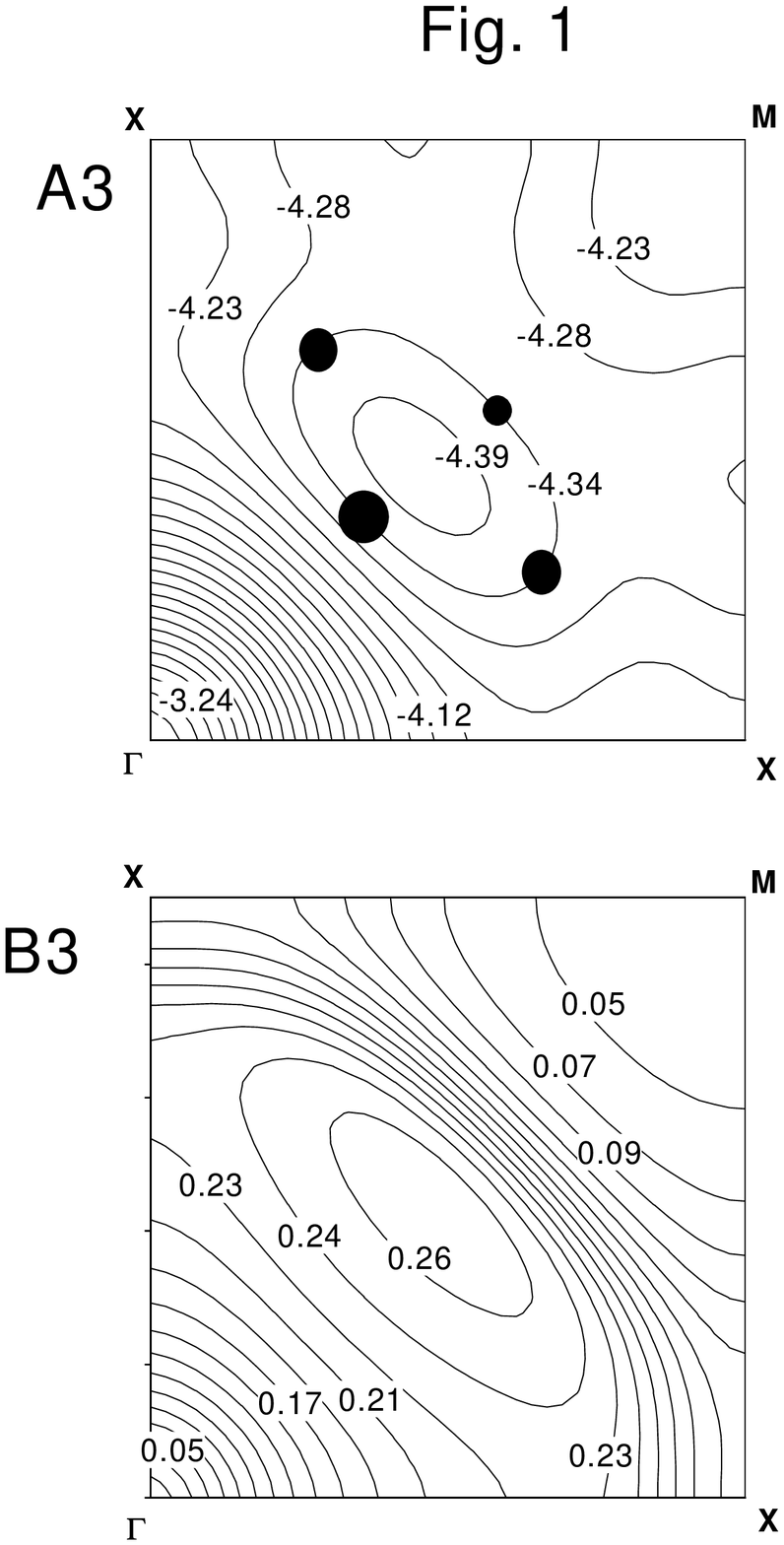,angle=0,width=5.3cm}}
\end{figure}
\end{minipage} \\
\begin{minipage}[t]{5.3cm}
\begin{figure}
\mbox{\psfig{file=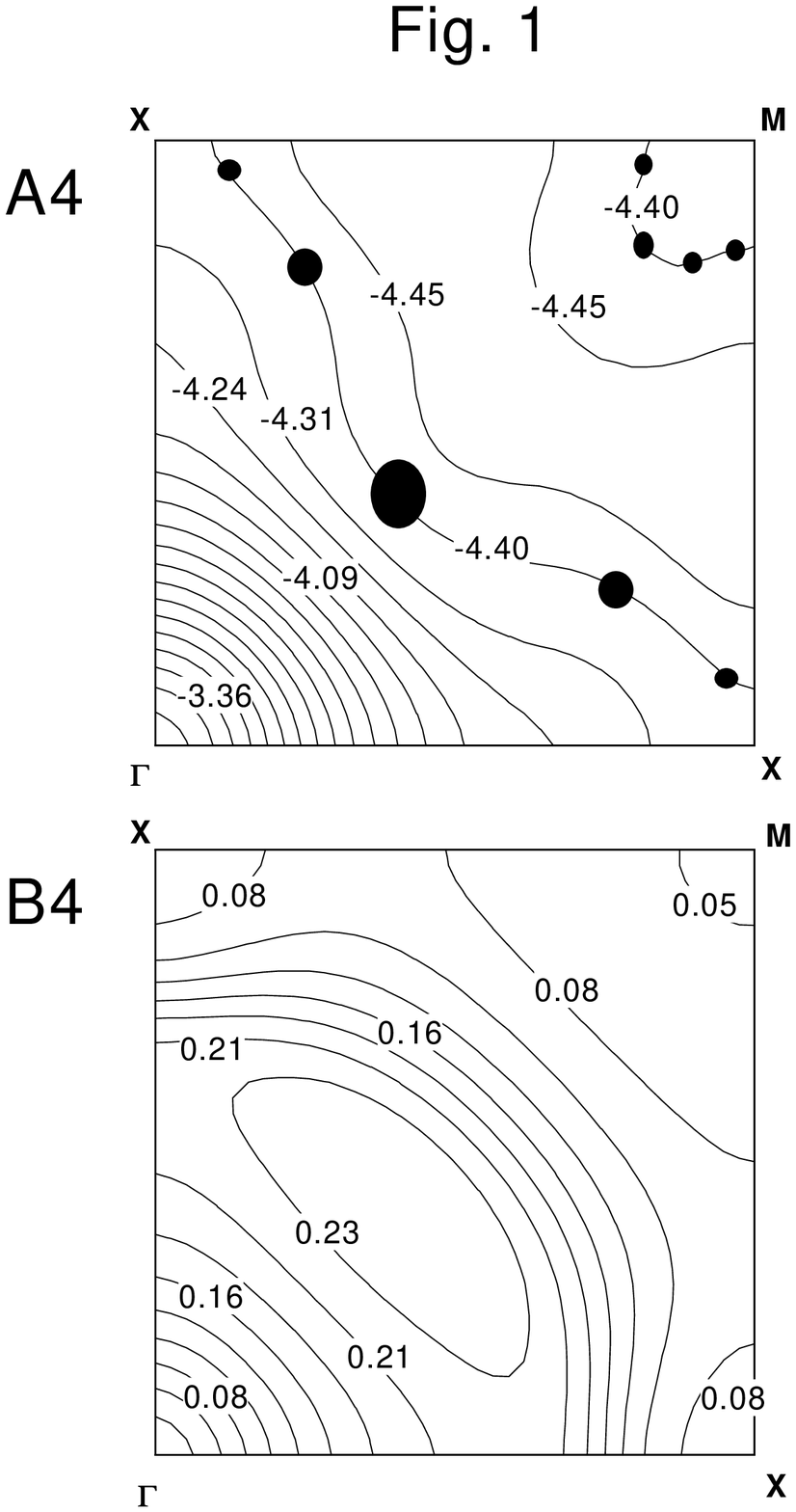,angle=0,width=5.3cm}}
\end{figure}
\end{minipage} 
\hspace*{-1cm}
\begin{minipage}[t]{5.3cm}
\begin{figure}
\mbox{\psfig{file=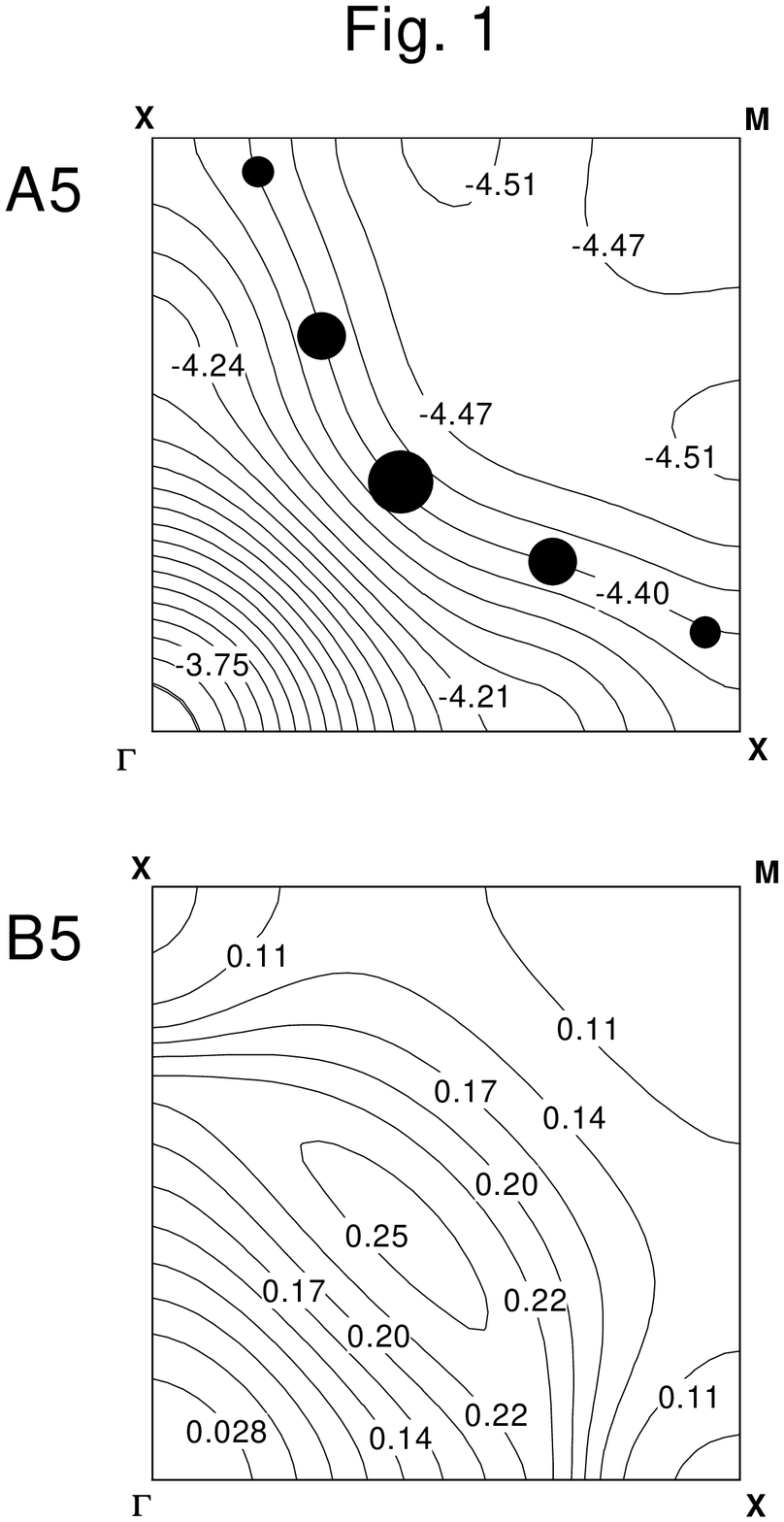,angle=0,width=5.3cm}}
\end{figure}
\end{minipage}
\hspace*{-1cm}
\begin{minipage}[t]{5.3cm}
\begin{figure}
\mbox{\psfig{file=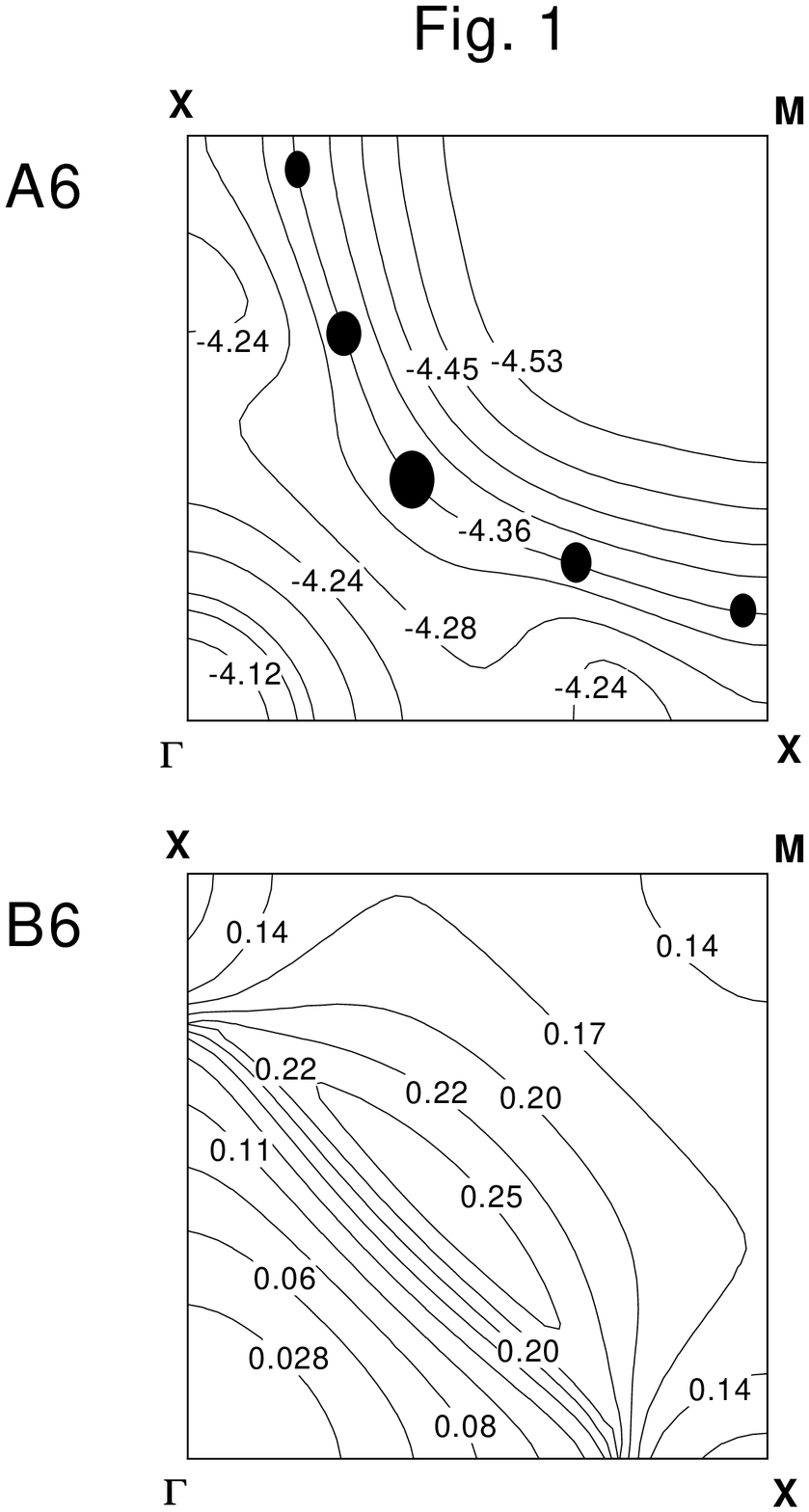,angle=0,width=5.3cm}}
\end{figure}
\end{minipage} \\
\begin{minipage}{16cm}
\baselineskip 5mm
{\bf Fig.1}: Spectrum $\varepsilon_1({\bf k})$ (figures a) and 
bare hole spectral weight $n_{h,\sigma }^{(1)}({\bf k})$  (figures b) for
the lowest band presented as contour plots 
(in units $\tau =1$) for different values of the frustration
parameter $p$ (0.05, 0.10, 0.13, 0.15, 0.20 and 0.25) corresponding to Figs.\ 
$1.1 \div 1.6$. The symmetry points in the first quadrant of the BZ are
denoted as $\Gamma
=(0,0),\quad X=\{(\pi ,0),(0,\pi )\},\quad M=(\pi ,\pi ).$
\end{minipage}
\end{minipage}
\vspace{2cm}

\noindent

\noindent
\begin{minipage}[t]{5.3cm}
\begin{figure}
\mbox{\psfig{file=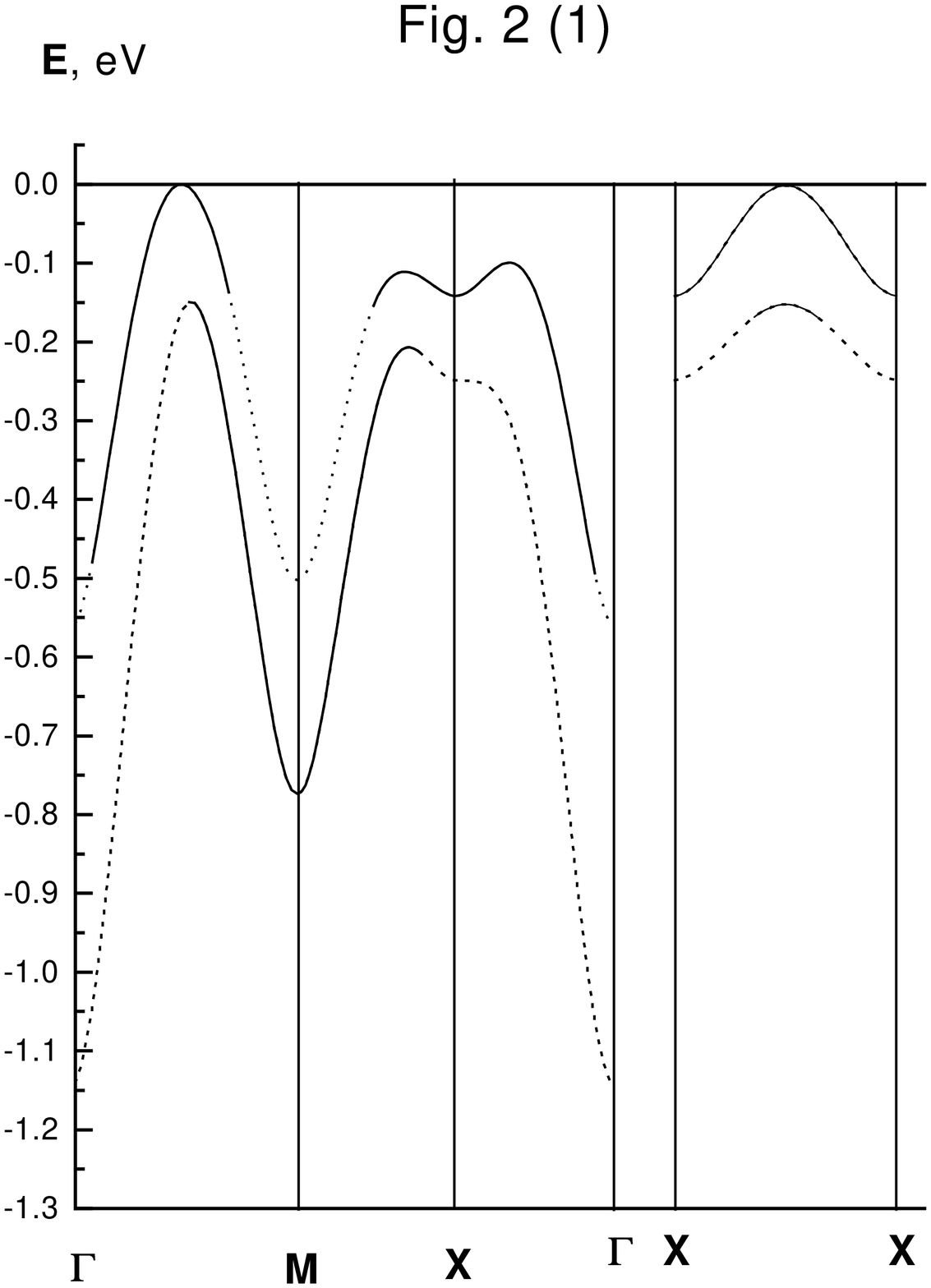,angle=0,width=5.3cm}}
\end{figure}
\end{minipage} 
\hspace*{-1cm}
\begin{minipage}[t]{5.3cm}
\begin{figure}
\mbox{\psfig{file=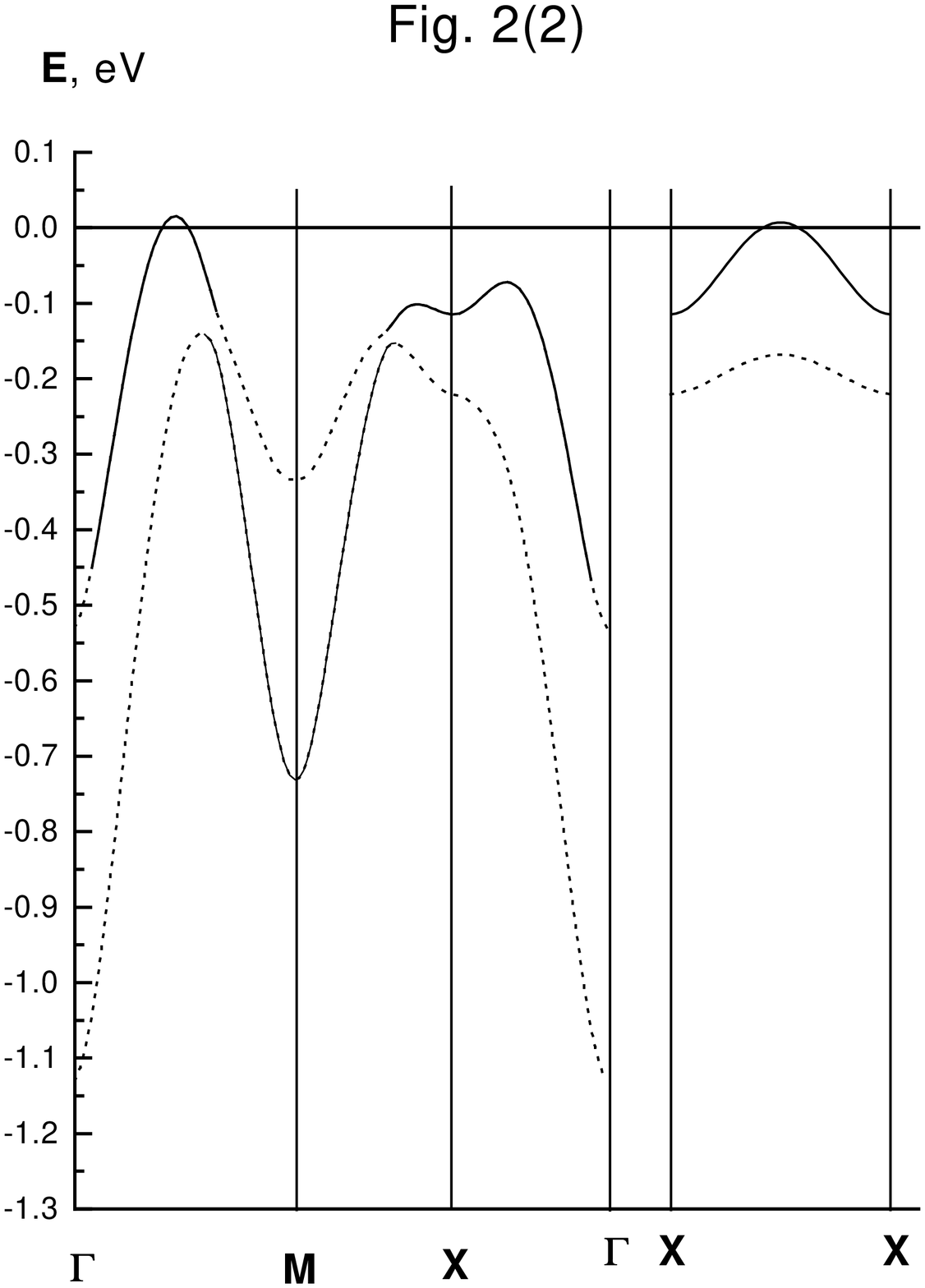,angle=0,width=5.3cm}}
\end{figure}
\end{minipage}
\hspace*{-1cm}
\begin{minipage}[t]{5.3cm}
\begin{figure}
\mbox{\psfig{file=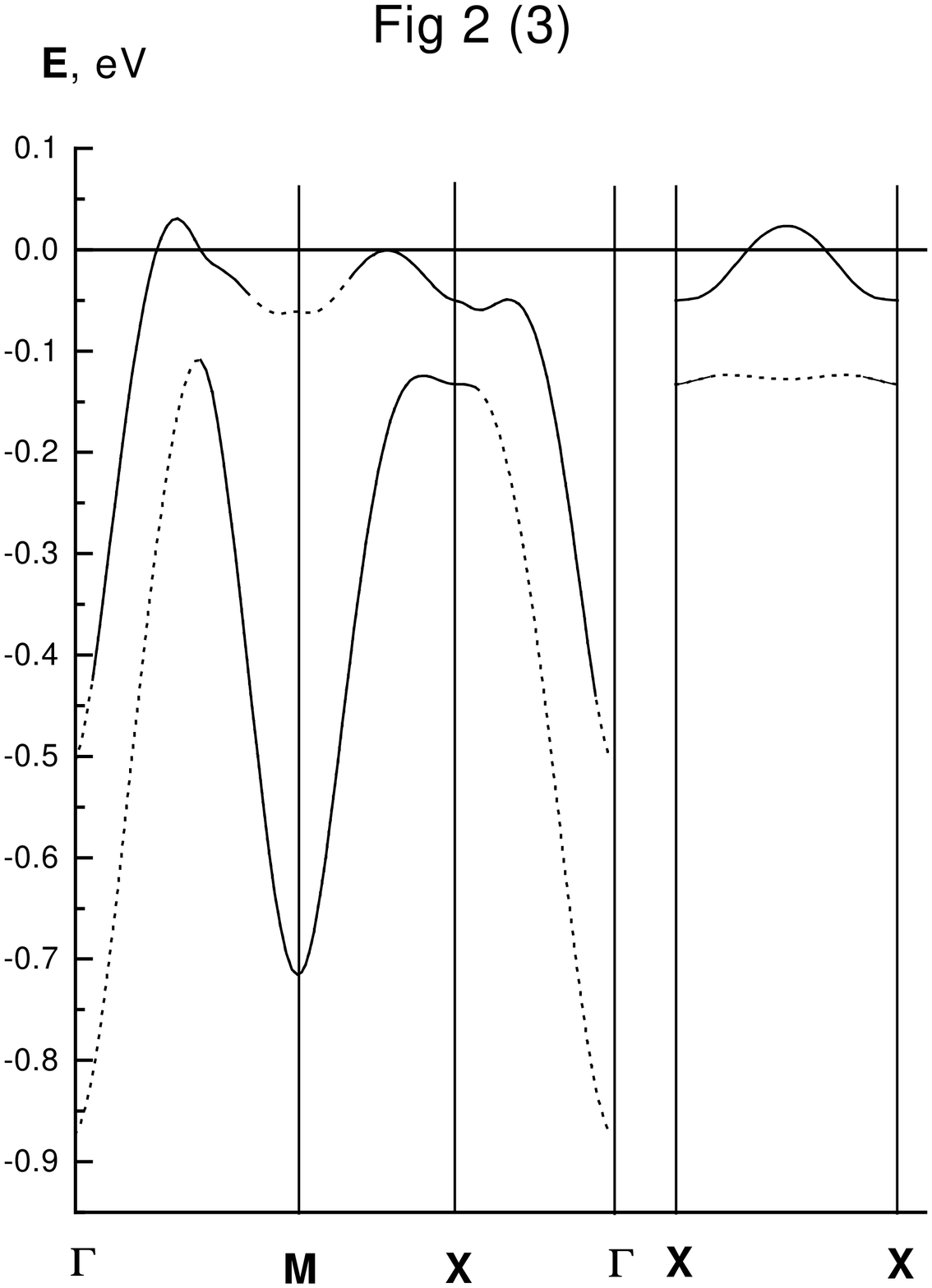,angle=0,width=5.3cm}}
\end{figure}
\end{minipage} \\
\begin{minipage}[t]{5.3cm}
\begin{figure}
\mbox{\psfig{file=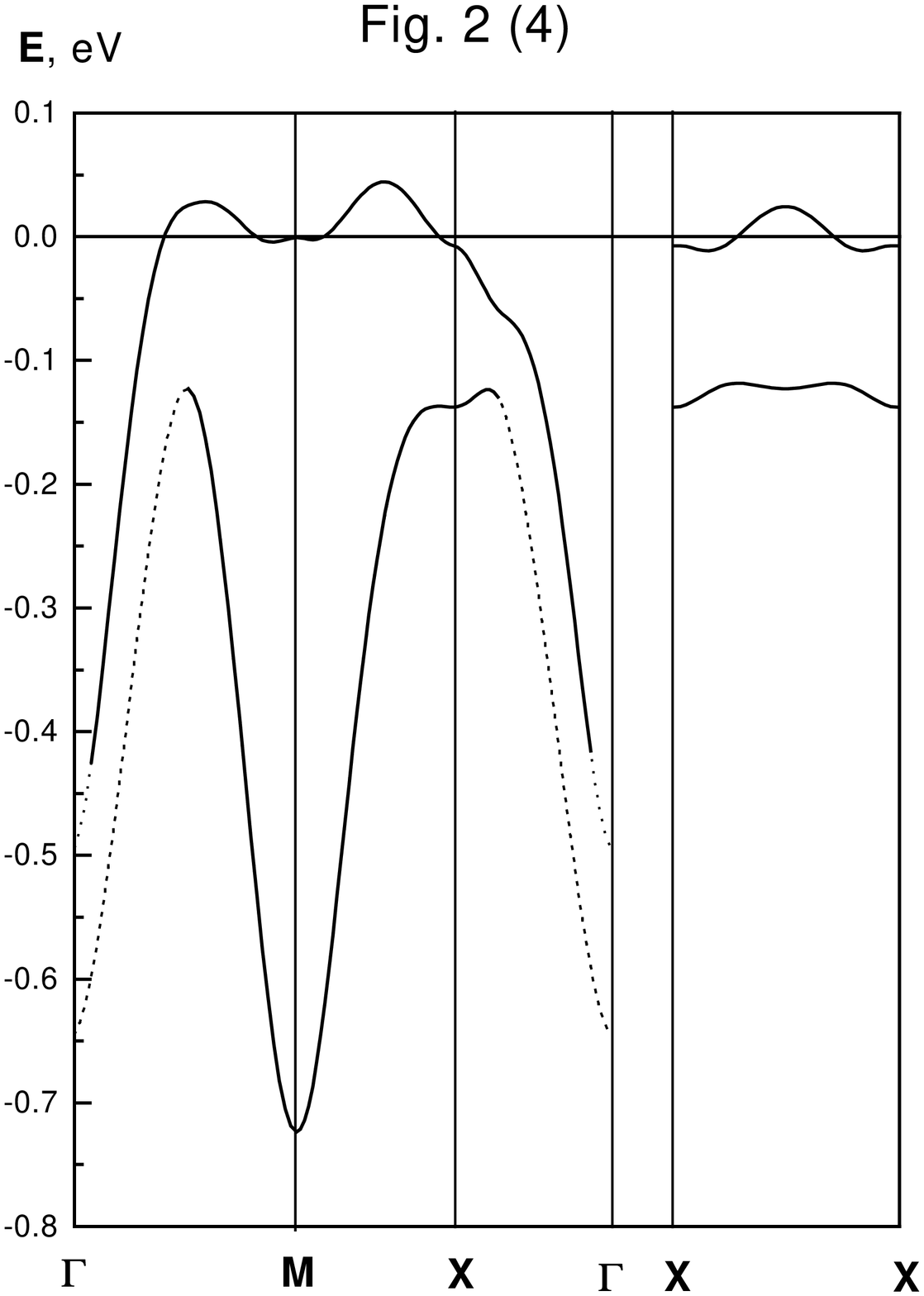,angle=0,width=5.3cm}}
\end{figure}
\end{minipage} 
\hspace*{-1cm}
\begin{minipage}[t]{5.3cm}
\begin{figure}
\mbox{\psfig{file=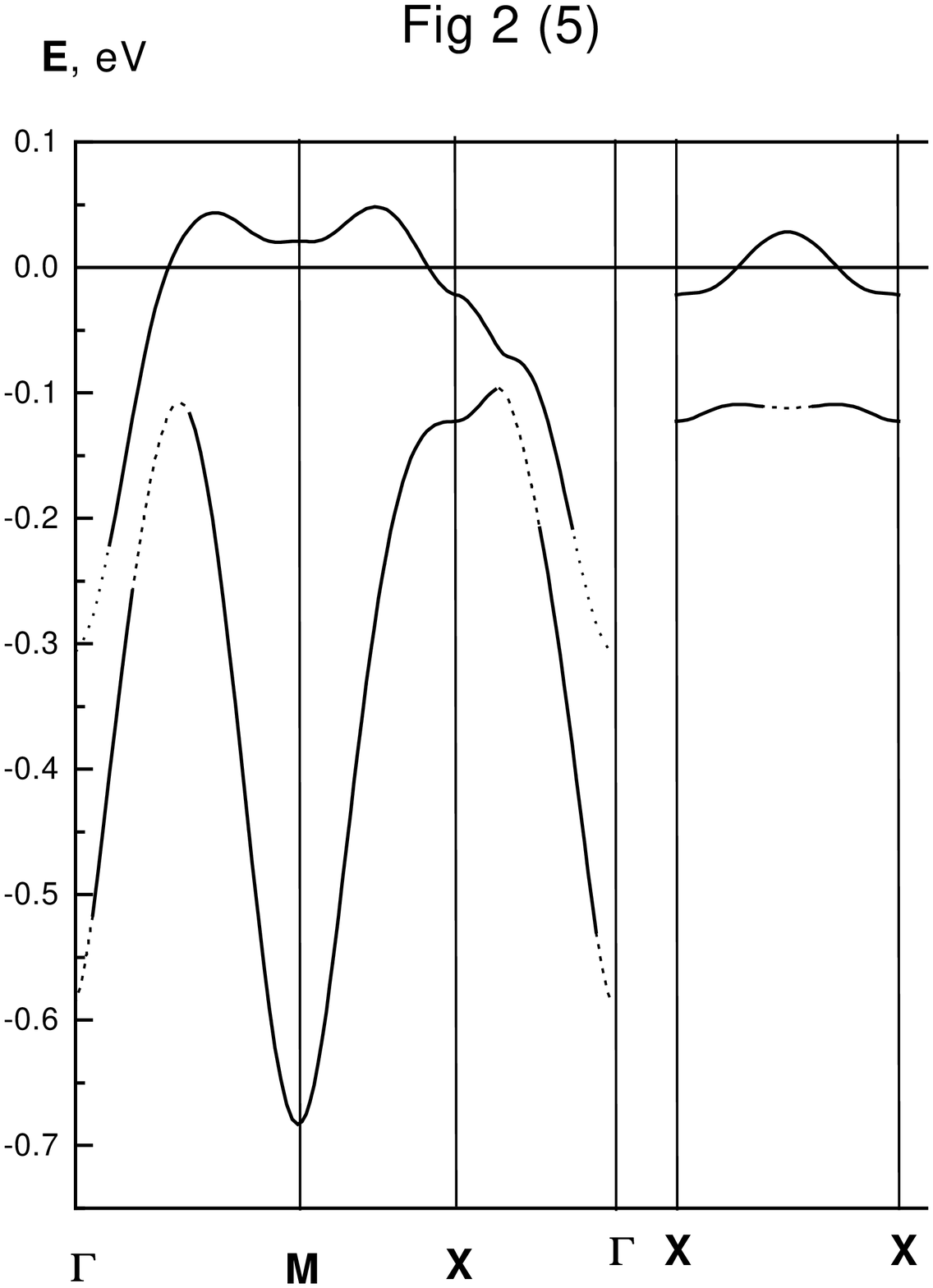,angle=0,width=5.3cm}}
\end{figure}
\end{minipage}
\hspace*{-1cm}
\begin{minipage}[t]{5.3cm}
\begin{figure}
\mbox{\psfig{file=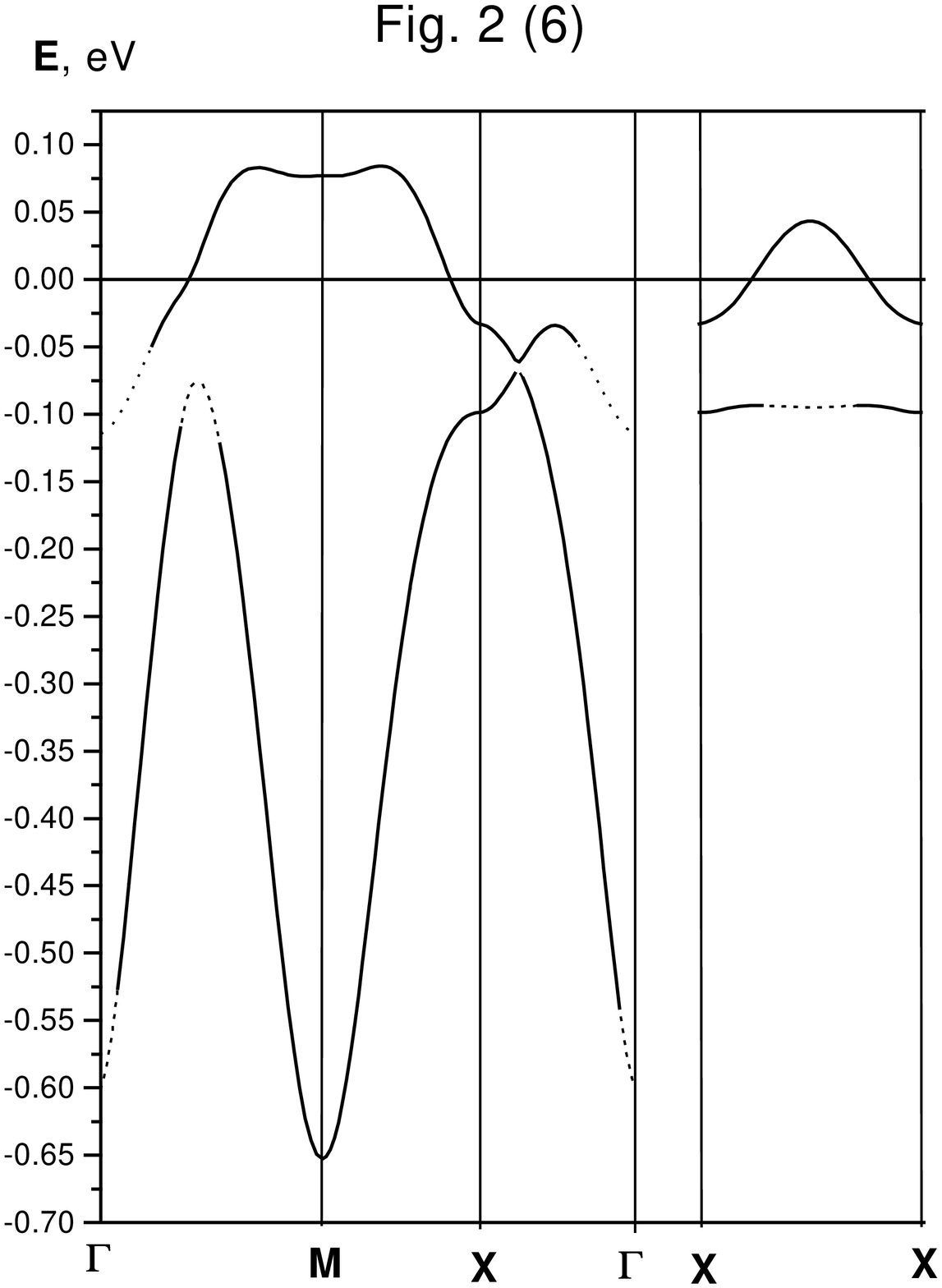,angle=0,width=5.3cm}}
\end{figure}
\end{minipage} \\

\noindent
{\bf Fig.2}: Electronic spectrum of the two lowest hole bands for the same
frustration parameters as in Fig.\ 1 along high symmetry lines of the BZ. Zero 
energy corresponds 
to the Fermi level and the value of $\tau$ has been set to $0.4$ eV. 
The parts of both bands which have considerable spectral
weight $n_{h,\sigma }^{(1,2)}({\bf k})>0.05$ are plotted by solid lines, the
parts with small weight by dotted lines.

\vspace*{2cm}

\noindent
The relation adopted below $x\Leftrightarrow p$ 
is given in Table 1. Of course this relation is
a phenomenological one and it may need some rescaling. Nevertheless, the main
conclusions are 
preserved if we admit the realistic assumption that the small and optimal
dopings correspond to $p\approx 0\div 0.1$ and $p\approx 0.1\div 0.25$, 
accordingly.
The contours decorated by full circles correspond to the Fermi surfaces  
calculated for the hole dopings $x$
from Table 1. The diameter of the circles is proportional to the spectral
weight $n_{h,\sigma }^{(1)}({\bf k})$ along the FS.
In Fig.\ 2 we show the electronic spectrum for the two lowest
bands along high symmetry lines for the 
value of $\tau =0.4$ eV as adopted in Sec.\ 2. 
The zero of energy corresponds to the Fermi
level. These parts of both bands which have considerable 
spectral weight $n_{h,\sigma }^{(1,2)}({\bf k})$ are plotted by solid lines,
those parts with small $n_{h,\sigma }^{(1,2)}({\bf k})<0.05$ by dotted lines.

Let us first discuss the dielectric (insulator) or heavily
underdoped regimes which are represented by Figs.\ 1.1, 1.2 and 2.1, 2.2.
In Figs.\ 1.1.a and 1.2.a the minimum of $\varepsilon _1({\bf k})$ is close to 
the $N$-point and the spectrum is rather isotropic near the band bottom. The
dispersions along the directions $\Gamma-M$, and $\Gamma-X-M$ reproduce
the ARPES results (compare for example Fig.\ 2.1 with the dispersion in Fig.\ 
3 of Ref. \cite{mdlpm96}). 
The band width of the first band $W_1\approx (4.4-3.3)\tau =0.44$ eV 
reflects also the experimental ARPES results  
$W_1\approx 0.2$ eV for Bi$_2$Sr$_2$CaCuO$_{6+0.5}$ 
\cite{mdlpm96}, $W_1\approx 0.3$ eV for Ca$_2$CuO$_2$Cl$_2$
\cite{rkfm98}, or $W_1\approx 0.35$ eV for La$_2$CuO$_4$ \cite{wsmkk95}. 
Some of the 
uncertainty in these experimental values might be related to the drop of 
spectral weight near the $\Gamma $-point.

The most important result in the heavily underdoped regime is the sudden drop
of spectral weight in the lowest band as ${\bf k}$ moves from $N$ to $M$, 
see Figs.\ 1.1.b and 1.2.b. The position of the ${\bf k}$-line where this 
drop occurs is close to that one which gives the remnant FS in the ARPES
experiment \cite{rkfm98}. The spin-polaron spectrum in Figs.\ 1.1.a 
and 1.2.a (2.1, 2.2) 
has a symmetry close to the symmetry of the magnetic BZ, but the
spectral weight $n_{h,\sigma }^{(1)}({\bf k})$ has the symmetry of the
initial BZ, see Figs. 1.1.b and  1.2.b.

For the case $p=0.25,\ x=0.19$,  Figs.\ 1.6., 2.6, which we treat as 
close to optimal doping, a large FS centered at $(\pi ,\pi )$ is seen. It
has a form consistent to the Luttinger theorem but with full filling $%
(1+x)$,  not $x$. The spectral weight map from Fig.1.6.b demonstrates
that such a large FS takes place due to small bare hole spectral weight 
at the 
spin-polaron ${\bf k}$-states under the FS. The average value of 
$n_{h,\sigma}^{(1)}({\bf k})$ is close to $0.17\ll 1$.  
Let us note that the average value
of $n_{h,\sigma }^{}({\bf k})$ for ${\bf k}$ below the FS would be 
approximately two times greater 
if we should take
the local polaron approach, i.e.\ restrict the basis operator set to 
(\ref{p1}). One would loose even a qualitative correspondence to the
experimental FS.

Let us compare the two lowest bands in the dielectric and the optimally doped
cases  
along the $N$ to $\Gamma$ cut. In the former case, see Fig.\ 2.1, only the
lowest hole polaron band is important (e.g. the upper electron one in 
Fig.\ 2.1), as the upper hole band has a small spectral weight. For the
optimally doped case, Fig.\ 2.6, the second band is important as well. The
correspondence to the ARPES experiment (see Fig. 3 in \cite{mdlpm96}) is
good if we assume that parts of the spectra are missed where the spectral
weight is small (dashed lines in Fig.2.6) and if we treat the second band in
the 
region close to $\Gamma $ as an 
extension of the first one. Such a treatment is correct if we take
into account the broadening of the bands. In our approach such a broadening
would be equivalent to an additional splitting of bands due to 
additional operators (relative to the set
(\ref{p7}), (\ref{p8})). Such an improvement 
was investigated for the unfrustrated regular Kondo
lattice model \cite{bkub00}. In the optimally doped case ($p=0.25$) the
effective bandwidth $W$ is $W=\varepsilon_{2,\max }-$ $\varepsilon
_F\approx 0.55$ eV, see Fig.\ 2.6, 
($W\approx 0.38$ eV for nearly optimally doped 
Bi$_2$Sr$_2$CaCuO$_{6+\delta}$ \cite{mdlpm96,gbl00,lbd00}). 
The comparison with the dielectric state demonstrates some band narrowing 
with doping reduction as it is also given by ARPES \cite{mdlpm96}.

An important feature of the spectrum is the presence of a second band $%
\varepsilon _2({\bf k})$ well along the $N$ to $M$ cut, see Fig.\ 2.6. This
branch resembles the main branch along the $N$ to $\Gamma $ cut if we shift
it in the ${\bf k}$-space by ${\bf Q}$. Relative to the main
branch along the $N$ to $M$ cut the band $\varepsilon _2({\bf k})$ is
also shifted in energy by $0.1\div 0.2$ eV. This branch is present as well 
in the
underdoped case, see Fig.\ 2.4. An analogous branch (but without energy shift)
was recently reported for underdoped $Bi2212$ \cite{bgl00} and it was
treated as the shadow band first observed in Ref.\ \cite{aos94}. For the
optimally doped compounds the presence of this band can be important 
for the plasma frequency at about $\omega_p=1$ eV if one takes into account the
corresponding interband transitions. Let us mention
that in the ordinary band theory such transitions start at
high energy $\approx 1.2$ eV and lead only to a constant
contribution in the static dielectric constant \cite{gpd99}. 

Our calculations reproduce also the extended saddle point
which is stretched in the direction from $X$ to $\Gamma$ close
to the FS (see Fig.\ 1.6.a). Such an extended saddle point is observed in the
photoemission data  
of optimally doped compounds and there has been a lot of speculations that it 
is 
important to the physics of the superconductors \cite{acg93}. Note that for
this flat band region (close to $(2\pi /3,0)$-point) the bare carrier
spectral weight $n_{h,\sigma }^{(1)}({\bf k})$ is not small and is close to $%
0.22,$ see Fig.1.6.b .

The underdoped case is represented in Figs.\ 1.3, 2.3 with $p=0.15$, $x=0.06$
and in Figs.\ 1.4, 2.4 ($p=0.15$, $x=0.11$). As we move from
doping 0.02 to doping 0.11 (compare in consecutive order Figs. 1.2.b,
1.3.b and 1.4.b), the topology of lines with equal spectral weight changes
dramatically. The remnant FS disappears, there appears finite spectral weight 
$n_{h,\sigma }^{(1)}({\bf k})$ close to the $M$ point, the curvatures of lines 
change sign in a broad region and the 
spectral weight along the cut $X-(\pi ,\pi /2)$ decreases. Between $X$ and
$(\pi,\pi/2)$ the spectral weight is transferred to the second band. Since
that is also the region where the FS crosses the $X$-$M$-line with increasing
doping, the $n_{h,\sigma }^{(1)}({\bf k})$ spectral weight along the FS is
strongly changed. Such a strong decrease of spectral weight as ${\bf k}$ moves 
along the FS from the central region around $N=(\pi/2,\pi/2)$ to points close
to $X=(\pi,0)$ can be treated as the opening of a high energy pseudogap
near the point $(\pi ,0)$.

This pseudogap $\delta $ is often determined as 
$\delta =\varepsilon (\pi,0)-\varepsilon _f$ (see \cite{rkfm98}). 
In our underdoped case ($p=0.15,\ x=0.11$) $n_{h,\sigma }^{(1)}(\pi ,0)$ 
is small and the first band 
should be only weakly present at the $X$ point in ARPES. So, we must treat 
$\varepsilon_2(\pi ,0)$ as $\varepsilon (\pi ,0)$ in the expression for the
gap $\delta$ (see Table 1). Then for
the adopted above value $\tau =0.4$ eV the value of the pseudogap is equal to 
$0.19$ eV in accordance with ARPES results ($0.1\div 0.2$ eV 
\cite{mdlpm96,lsd96,dyc96,dnytr97}). The dependence of the 
spectral weight $n_{h,\sigma }^{(1)}(\pi ,0)$ on doping given in Table 1
is non monotonic. At first, a strong drop of $n_{h,\sigma }^{(1)}(\pi ,0)$
takes 
place in the underdoped region, but the spectral weight is sufficiently large
in the optimally doped regime. Such a behavior reflects the ARPES scenario for 
a decreasing pseudogap as the doping increases from $0.1$ to $0.2$.

The evolution of the lowest band dispersion shows that in the just mentioned
transition region also the formation of the extended saddle point takes place
close to the $(2\pi /3,0)$-point. As the doping increases above $x=0.11$ this
flat band region is preserved. 

For any doping, the particle spectral weight $n_{h,\sigma }^{(1)}({\bf k})$
for ${\bf k}$ close to the $N$ point is large and nearly constant 
$n_{h,\sigma }^{(1)}({\bf k})=0.22\div 0.25$, see Table 1. 
And indeed, in this ${\bf k}$-region the ARPES experiments demonstrate the
position of the FS surface with a 
well pronounced quasiparticle peak. Accordingly,  the second band in this 
${\bf k}$-region gives a small value $n_{h,\sigma }^{(2)}({\bf k})<0.03$
(a small incoherent part) at any doping.

\vspace*{1cm}

\noindent
{\bf Table 1:} Frustration parameter $p$, doping $x$, spin correlation length 
$\xi_{\bf Q}$ and cut-off momentum $\kappa$ for which calculations have been
done. The derived energies ($\varepsilon^{(1)}(\pi,0)$, 
$\varepsilon^{(2)}(\pi,0)$, and $\varepsilon_F$) are given in units of
$\tau$. $n_{h,\sigma }^{(1,2)}({\bf k})$ denotes the spectral weight of the
first or second band, respectively. 

\vspace*{0.5cm}

$
\begin{array}{|c|c|c|c|c|c|c|c|c|c|}
\hline
p & \kappa _0 & \xi _{{\bf Q}} & x & n_{h,\sigma }^{(1)}(\pi ,0) & 
n_{h,\sigma }^{(2)}(\pi ,0) & n_{h,\sigma }^{(1)}{\bf (}\pi /2,\pi /2{\bf )}
& \varepsilon ^{(1)}(\pi {\bf ,}0{\bf )} & \varepsilon ^{(2)}(\pi {\bf ,}0%
{\bf )} & \varepsilon _F \\ 
\hline
0.05 & 0.08\pi  & 12 & 0.00 & 0.23 & 0.02 & 0.23 & -4.06 &  & (-4.42) \\ 
0.10 & 0.17\pi  & 6  & 0.02 & 0.25 & 0.00 & 0.26 & -4.09 &  & -4.38 \\ 
0.13 & 0.30\pi  & 3.4& 0.06 & 0.13 & 0.11 & 0.27 & -4.21 & -4.04 & -4.34 \\ 
0.15 & 0.35\pi  & <3 & 0.11 & 0.06 & 0.18 & 0.23 & -4.38 & -4.03 & -4.40 \\ 
0.2 & 0.35\pi  & <2  & 0.14 & 0.08 & 0.16 & 0.24 & -4.35 & -4.08 & -4.40 \\ 
0.25 & 0.35\pi  & <1 & 0.19 & 0.10 & 0.13 & 0.25 & -4.28 & -4.10 & -4.36 \\
\hline
\end{array}
$

\section{CONCLUSION}

We presented a semi-phenomenological approach to describe the evolution of
Fermi surface and electronic structure with doping. The approach is based on
the spin-fermion model for the CuO$_2$ plane and simulates the doping by a
frustration term in the spin Hamiltonian. It shows a strong alteration of the
electronic structure which is due to a simple and natural
mechanism: doping leads to frustration in the spin subsystem and to the
variation of spin-spin correlation functions, and correspondingly to 
the non-rigid behavior of the spectrum.

Our theory explains the isotropic band bottom, the large energy difference
between $N$ and $X$ \cite{wsmkk95,mdlpm96} and the remnant FS \cite{rkfm98} of 
the undoped compounds. In the optimally doped case it shows a flat band region 
(an extended saddle point) between $(\pi/2,0)$ and $(\pi,0)$, a large FS
\cite{togls92,gcdgl93,acg93,dskml93,ksdmp94} and it shows signatures of a
shadow FS \cite{aos94,bgl00}. We find in the underdoped region 
a rather strong change of dispersion and spectral weight at the FS near
$X$. This finding can also be interpreted in terms of a pseudogap
\cite{lsd96,dyc96,dnytr97}. 

Our theoretical results would suggest for very low doping a FS in the form of
small hole pockets (see e.g.\ Fig.\ 1.3.a). However, the spectral weights at
those parts of the FS which are close to $M$ are rather small. This might be a
reason which hinders its observation in ARPES. In underdoped samples we find a 
transition to a large FS but with very small spectral weight near
$X=(\pi,0)$. Accordingly, we would interpret the Fermi surface arcs
\cite{ndr98} as those parts of the FS with large spectral weight. We find the
parts of the FS with large spectral weight near to $N$ to be remarkable stable 
with doping despite the tremendous changes in other parts of the BZ. 

There are several possible improvements of the approach presented here. A
straightforward way consists
in introducing of several non overlapping domains 
${\bf \Omega }_1,{\bf \Omega}_2,{\bf \Omega }_3,...{\bf \Omega }_i$
instead of the single domain ${\bf \Omega}$ (see also \cite{bkub00}). 
As a result such an extension of the operator set should lead to a better
description of the incoherent part of the spectral function. It might also
lead to a better description of the shadow FS and to a more abrupt change of
the spectral density along the FS in the underdoped case, i.e.\ a better
description of the pseudogap.

\vspace*{1cm}

\noindent
{\bf ACKNOWLEDGMENTS} \\
This work was supported, in part, by the INTAS\ (project No. 97-11066) and
by RFFS (projects No. 98-02-17187).


\vspace*{2.5cm}

\baselineskip 5mm

\noindent
\Large
{\bf Appendix A} \\
\normalsize
\baselineskip 5mm
Let us present the explicit form of Green's function $G_s({\bf k},\omega)$ 
and spin-wave spectrum $\omega ({\bf k)}$ for three degenerate modes 
in the framework of the spherically symmetric approach \cite{bb94a}: 

\begin{equation}
\label{a1}G_s({\bf k},\omega )=<S_{{\bf k}}^\alpha {\bf |}S_{-{\bf k}%
}^\alpha >=\frac{F({\bf k)}}{\omega ^2-\omega ^2({\bf k})},\quad \alpha
=x,y,z 
\end{equation}

\begin{equation}
\label{a2}F({\bf k)=-}2\left[ J_1z_gC_g(1-\gamma _g({\bf k}%
))+J_2z_dC_d(1-\gamma _d({\bf k}))\right] 
\end{equation}
\begin{equation}
\label{a3}
\begin{array}{l}
\omega ^2( 
{\bf k)=}\frac 23{\bf [}\left( 1-\gamma _g\right) \left(
J_1J_2K_{gd}+I_1^2(z_g(z_g-1)C_g\alpha _{_1}+\frac 34z_g+K_{gg})\right) \\ 
+ \left( 1-\gamma _d\right) \left( J_1J_2K_{gd}+J_2^2(z_d(z_d-1)C_d\alpha
_{_3}+\frac 34z_d+K_{dd})\right)  \\ 
-\left( 1-\gamma _g^2\right) J_1^2z_g^2C_g\alpha _{_1}-\left( 1-\gamma
_d^2\right) J_2^2z_d^2C_d\alpha _{_3} \\
-\left( 1-\gamma _g\right) \gamma_dJ_1J_2Z_gZ_dC_g\alpha_{_1} 
-\left( 1-\gamma _d\right) \gamma _gJ_1J_2Z_gZ_dC_d\alpha _{_3}] 
\end{array}
\end{equation}

$$
K_{gd}=\sum_{r=g+d} \alpha_rC_r \; ; \quad 
K_{gg}=\sum_{r=g_1+g_2}^{g_1\neq -g_2} \alpha _rC_r \; ;
K_{dd}=\sum_{r=d_1+d_2}^{d_1\neq -d_2} \alpha _rC_r \; .  
$$
Here $z_g,z_d$ are a number of first and second nearest neighbors on square
lattice and $\alpha _r$ are vertex corrections.

$\alpha _g=\alpha _{_1};\alpha _d=\alpha _{_3};\alpha _r=\alpha _{_2}$ if $%
r>d;$

$\frac{\alpha _{_1}-1}{\alpha _{_2}-1}=R_\alpha =0.863;$

$\alpha _{_3}=(1-p)\alpha _{_2}+p\alpha _{_1};$

$J_1=(1-p)J,J_2=pJ;$

$K_{gd}=8\alpha _{_1}C_g+8\alpha _{_2}C_f;K_{gg}=4\alpha _{_2}C_{2g}+8\alpha
_{_3}C_d;$

$K_{dd}=8\alpha _{_2}C_{2g}+4\alpha _{_2}C_{2d};$

\noindent
The values of $\alpha _{_1}$, $C_g$, $C_d$, $C_{2g}$, $C_{gd}$, $C_{2d}$ are
calculated self-consistently for each values of $p$ and $T$ \cite{bb94a}.
The equations (\ref{a1},\ref{a2},\ref{a3}) give an expression for the 
spin-spin structure factor $C({\bf k})$: 
\begin{equation}
\label{a4}C({\bf k})=A(1+e^{\omega _{{\bf k}}/T})/(e^{\omega _{{\bf k}%
}/T}-1),\quad A=F({\bf k)/}2\omega _{{\bf k}}. 
\end{equation}
The correlation length $\xi _{{\bf Q}}$ (relative to the N\'eel type phase) 
is determined by power expansion of the Green's function (\ref{a1}) on the
value ${\bf q}={\bf Q}-{\bf k}$ at $\omega =0$ \cite{shim93}:

\begin{equation}
\label{a5}G({\bf q},0)=G({\bf Q},0)/(1+\xi _{{\bf Q}}^2{\bf q}^2) 
\end{equation}

\noindent
\Large
{\bf Appendix B} \\
\normalsize
\baselineskip 5mm
Here we present the explicit form of $K$ and $D$ matrix elements.
The following notations are used below.

\medskip
\hangindent=1cm \noindent
$D_{ij}({\bf k})=\langle \{ [ A_{{\bf k},i}, (\hat \tau+\hat J+ \hat h) ],
A^+_{{\bf k},j} \} \rangle = \tau \tilde \tau _{ij}+J_1\tilde
J_{ij}^{(1)}+J_2\tilde J_{ij}^{(2)}- h\tilde h_{ij};$

\medskip
\hangindent=1cm \noindent
$K_{ij}, \tilde \tau _{ij}, \tilde J_{ij}^{(1)}, \tilde J_{ij}^{(2)}, \tilde
h_{ij}$ \quad are symmetric matrices.

\medskip
\hangindent=1cm \noindent
$\gamma _g=0.5(\cos k_x+\cos k_y);\gamma _d=\cos k_x\cos k_y;\gamma
_{2g}=0.5(\cos 2k_x+\cos 2k_y);$

\medskip
\hangindent=1cm \noindent
The nonzero matrix elements are as follows:

\bigskip
\hangindent=1cm \noindent
${\bf K-matrix}$

\smallskip
\hangindent=1cm \noindent
$K_{11}=0.5(1+\cos k_x);$

\hangindent=1cm \noindent
$K_{16}=0.5u(1+\cos k_x);$

\hangindent=1cm \noindent
$K_{22}=0.5(1+\cos k_y);$

\hangindent=1cm \noindent
$K_{26}=0.5u(1+\cos k_y);$

\hangindent=1cm \noindent
$K_{33}=3/4+C_g\gamma _g;$

\hangindent=1cm \noindent
$K_{34}=0.5(u+u_g\cos k_x);$

\hangindent=1cm \noindent
$K_{35}=0.5(u+u_g\cos k_y);$

\hangindent=1cm \noindent
$K_{36}=-u+\gamma _gC_g(v_g-v);$

\hangindent=1cm \noindent
$K_{44}=0.5(u+u_g\cos k_x);$

\hangindent=1cm \noindent
$K_{46}=-uv+0.5w+0.5\cos k_x(-v_gu-vu_g+w_g);$

\hangindent=1cm \noindent
$K_{55}=0.5(u+u_g\cos k_y);$

\hangindent=1cm \noindent
$K_{56}=-uv+0.5w+0.5\cos k_y(-v_gu-vu_g+w_g);$

\hangindent=1cm \noindent
$K_{66}=(3/4)u+2uv-w+\gamma _g(u_gC_g+u^2-(1/3)u_g^2-4C_g((1/3)u_gv+uv_g)$

$+2vv_gC_g+2v_g^2C_g^2+C_gv^2(1+(2/3)C_g))-(2/3)\gamma _gW_g^{(\tau )};$

\bigskip

\vspace*{2cm}

\hangindent=1cm \noindent
$\widetilde{{\bf \tau }}{\bf -matrix}$

\smallskip
\hangindent=1cm \noindent
$\tau _{11}=0.5(1+\cos k_x)^2;$

\hangindent=1cm \noindent
$\tau _{12}=0.5+\gamma _g+0.5\gamma _d;$

\hangindent=1cm \noindent
$\tau _{13}=(3/2+2C_g\gamma _g)(1+\cos k_x);$

\hangindent=1cm \noindent
$\tau _{14}=(1+\cos k_x)(u+u_g\cos k_x);$

\hangindent=1cm \noindent
$\tau _{15}=u(1+\cos k_x)+u_g(\cos k_y+\gamma _d);$

\hangindent=1cm \noindent
$\tau _{16}=(1+\cos k_x)(u(-1+\gamma _g)+2C_g\gamma _g(v_g-v));$

\hangindent=1cm \noindent
$\tau _{22}=0.5(1+\cos k_y)^2;$

\hangindent=1cm \noindent
$\tau _{23}=(1+\cos k_y)(3/2+2C_g\gamma _g);$

\hangindent=1cm \noindent
$\tau _{24}=u(1+\cos k_y)+u_g(\cos k_x+\gamma _d);$

\hangindent=1cm \noindent
$\tau _{25}=(1+\cos k_y)(u+u_g\cos k_y);$

\hangindent=1cm \noindent
$\tau _{26}=(1+\cos k_y)(u(-1+\gamma _g)+2C_g\gamma _g(v_g-v));$

\hangindent=1cm \noindent
$\tau _{33}=-9/8-4C_g\gamma _g+C_g+0.5C_{2g}\gamma _{2g}+C_d\gamma _d; $

\hangindent=1cm \noindent
$\tau _{34}=-(3/4)u-0.5vC_g-u_g\cos k_x+(\cos^2 k_x-0.5)
(v_{2g}C_g+0.5u_{2g}) $

$+\gamma _g(u_g+2v_gC_g-2v_gC_g\cos k_x-2vC_g)+\gamma _d(0.5u_d+v_dC_g);$

\hangindent=1cm \noindent
$\tau _{35}=-(3/4)u-0.5vC_g-u_g\cos k_y+(\cos^2 k_y-0.5)
(v_{2g}C_g+0.5u_{2g}) $

$+\gamma _g(u_g+2v_gC_g-2v_gC_g\cos k_y-2vC_g)+\gamma _d(0.5u_d+v_dC_g);$

\hangindent=1cm \noindent
$\tau _{36}=(9/2)u-(1/4)u_g+\gamma _g(3u+4vC_g+3u_g-4v_gC_g)$

$+\gamma _{2g}(C_gu+0.5v_{2g}C_{2g}-(2/3)v_{2g}C_gC_{2g}-0.5vC_{2g}$

$-(2/3)vC_gC_{2g}-(1/3)u_gC_{2g}-2v_gC_g^2+u_{2g}C_g)$

$+\gamma
_d(2C_gu+v_dC_d-vC_d-(4/3)C_gC_d(v_d+v)-(2/3)u_gC_d-4v_gC_g^2+2u_dC_g);$

\hangindent=1cm \noindent
$\tau _{44}=(3/4)u-v_gu_g-2uv+w+0.5w_g^{(1)}+\cos k_x(u_g-2vu_g-2v_gu+2w_g)$

$+(\cos^2 k_x-0.5)((1/2)u_{2g}-2v_gu_g+w_g^{(2)});$

\hangindent=1cm \noindent
$\tau _{45} =0.5u+2\gamma _g(-vu_g-v_gu+w_g+0.5u_g)-2uv+w+\gamma
_d(0.5u_d-2u_gv_g+w_g^{(3)}); $

\hangindent=1cm \noindent
$\tau _{46}=1.5uv+1.5u-0.75w+0.5(uC_g+C_gv^2+C_gvv_g+C_gv_g^2-(8/3)u_gv_gC_g$

$+(2/3)uu_g-(8/3)uvC_g+(8/3)vv_gC_g^2)-(1/3)W_1^{(\tau )}$

$+\gamma
_g(-v_gu-vu_g+w_g+2u_gC_g+2v^2C_g+(4/3)v^2C_g^2+4vv_gC_g+4v_g^2C_g^2 $

$+2u^2-(2/3)u_g^2-(8/3)u_gvC_g-8uv_gC_g-(4/3)W_g^{(\tau )})$

$+\cos k_x(-w_g+1.5u_g+uv_g+u_gv)$

$%
+(2cos^2k_x-1)(0.25w_{2g}-0.25uv_{2g}-0.25vu_{2g}+0.5C_gu_{2g}+0.5C_gvv_{2g} 
$

$%
+C_g^2v_gv_{2g}-(1/3)C_gu_{2g}v-C_guv_{2g}-(1/6)u_gu_{2g}+0.5C_gvv_g+0.5C_gv_g^2 
$

$+(1/3)C_g^2vv_g-(4/3)C_gu_gv_g+0.5uu_g-(1/3)W_2^{(\tau )})$

$+2\gamma
_d(0.25w_d-0.25uv_d-0.25vu_d+0.5C_gu_d+0.5C_gvv_d+C_g^2v_gv_d-(1/3)C_gu_dv$

$-C_guv_d-(1/6)u_gu_d+0.5C_gvv_g+0.5C_gv_g^2+(1/3)C_g^2vv_g-(4/3)C_gu_gv_g$

$+0.5uu_g-(1/3)W_3^{(\tau )});$

\hangindent=1cm \noindent
$\tau _{55}=(3/4)u-v_gu_g-2uv+w+0.5w_g^{(1)}+\cos
k_y(u_g-2vu_{2g}-2v_gu+2w_g)$

$+(\cos ^2k_y-0.5)((1/2)u_{2g}-2v_gu_g+w_g^{(2)});$

\hangindent=1cm \noindent
$\tau _{56}=1.5uv+1.5u-0.75w+0.5(uC_g+C_gv^2+C_gvv_g+C_gv_g^2-(8/3)u_gv_gC_g$

$+(2/3)uu_g-(8/3)uvC_g+(8/3)vv_gC_g^2)-(1/3)W_1^{(\tau )}$

$+\gamma
_g(-v_gu-vu_g+w_g+2u_gC_g+2v^2C_g+(4/3)v^2C_g^2+4vv_gC_g+4v_g^2C_g^2 $

$+2u^2-(2/3)u_g^2-(8/3)u_gvC_g-8uv_gC_g-(4/3)W_g^{(\tau )})$

$+\cos k_y(-w_g+1.5u_g+uv_g+u_gv)$

$+(2\cos
^2k_y-1)(0.25w_{2g}-0.25uv_{2g}-0.25vu_{2g}+0.5C_gu_{2g}+0.5C_gvv_{2g}$

$%
+C_g^2v_gv_{2g}-(1/3)C_gu_{2g}v-C_guv_{2g}-(1/6)u_gu_{2g}+0.5C_gvv_g+0.5C_gv_g^2 
$

$+(1/3)C_g^2vv_g-(4/3)C_gu_gv_g+0.5uu_g-(1/3)W_2^{(\tau )})$

$+2\gamma
_d(0.25w_d-0.25uv_d-0.25vu_d+0.5C_gu_d+0.5C_gvv_d+C_g^2v_gv_d-(1/3)C_gu_dv$

$-C_guv_d-(1/6)u_gu_d+0.5C_gvv_g+0.5C_gv_g^2+(1/3)C_g^2vv_g-(4/3)C_gu_gv_g$

$+0.5uu_g-(1/3)W_3^{(\tau )});$

\hangindent=1cm \noindent
$\tau _{66}=-9vu-(9/8)u+(9/2)w-(1/4)w_g^{(1)}+u(C_g+(2/3)u_g-(8/3)C_gv)$

$+v^2C_g+v_g^2C_g+vv_g(C_g+(8/3)C_g^2)+v_g(0.5u_g-(8/3)C_gu_g)-(2/3)W_1^{(%
\tau )}$

$+\gamma
_g(vu_g(-6+(16/3)C_g)+v_gu(-6+16C_g)-8vv_gC_g-v^2(4C_g+(8/3)C_g^2)-4u^2$

$+(4/3)u_g^2-8v_g^2C_g^2-4u_gC_g+6w_g+(8/3)W_g^{(\tau )})$

$+\gamma
_{2g}(0.5u_{2g}C_{2g}+vv_{2g}(C_{2g}+(4/3)C_gC_{2g})+v^2(0.5C_{2g}+(4/3)C_gC_{2g}+(1/3)C_{2g}^2) 
$

$%
+v_{2g}^2(C_{2g}^2+(8/3)C_gC_{2g})+v(-(2/3)u_{2g}C_{2g}+(2/3)u_gC_{2g}-2u_{2g}C_g-2uC_g) 
$

$%
-v_{2g}(2uC_{2g}+(2/3)u_gC_{2g}+(2/3)u_{2g}C_g+2uC_g)+0.5u^2-(1/6)M_{2g}^2-(14/3)v_g^2C_g^2 
$

$%
+vv_g(4C_g^2-(4/3)C_gC_{2g})+v_gv_{2g}(8/3)(C_g^2-C_gC_{2g})+v_g((2/3)u_gC_{2g}+(2/3)u_{2g}C_g 
$

$+2uC_g)+2w_{2g}C_g-(1/3)w_g^{(2)}C_{2g}-(1/3)W_{2g}^{(\tau )})$

$+2\gamma
_d(0.5u_dC_d+vv_d(C_d+(4/3)C_gC_d)+v^2(0.5C_d+(4/3)C_gC_d+(1/3)C_d^2)$

$+v_d^2(C_d^2+(8/3)C_gC_d)+v(-(2/3)u_dC_d+(2/3)u_gC_d-2u_dC_g-2uC_g)$

$%
-v_d(2uC_d+(2/3)u_gC_d+(2/3)u_dC_g+2uC_g)+0.5u^2-(1/6)u_d^2-(14/3)v_g^2C_g^2 
$

$%
+vv_g(4C_g^2-(4/3)C_gC_d)+v_gv_d(8/3)(C_g^2-C_gC_d)+v_g((2/3)u_gC_d+(2/3)u_dC_g 
$

$+2uC_g)+2w_dC_g-(1/3)w_g^{(3)}C_d-(1/3)W_d^{(\tau )});$

\bigskip
\hangindent=1cm \noindent
$\widetilde{{\bf J}}^{(1)}{\bf -matrix}$

\smallskip
\hangindent=1cm \noindent
$J_{33}=-4C_g+C_g\gamma _g;$

\hangindent=1cm \noindent
$J_{34}=2C_g(v_g-v)+\cos k_xC_g(0.5v-2v_g+0.5v_{2g}+v_d);$

\hangindent=1cm \noindent
$J_{35}=2C_g(v_g-v)+\cos k_yC_g(0.5v-2v_g+0.5v_{2g}+v_d);$

\hangindent=1cm \noindent
$J_{36}=2C_g(v-v_g)+2u_g+\gamma
_g(0.5C_g(v_g-v)-2C_g^2v_g-0.5u_g-(2/3)u_gC_{2g}$

$%
-(4/3)u_gC_d+(4/3)v_dC_gC_d+(2/3)v_{2g}C_gC_{2g}+(2/3)u_{2g}C_g+(4/3)u_dC_g); 
$

\hangindent=1cm \noindent
$J_{44}=2C_g(v_g-v)+C_g\cos k_x(0.5v-2v_g+0.5v_{2g}+v_d);$

\hangindent=1cm \noindent
$%
J_{46}=(4/3)C_g(w_g-w)+(4/3)C_gu(v-v_g)+v_g^2(C_g+(4/3)C_g^2)+2v^2C_g+(4/3)W_{g4}^{(J)} 
$

$+vv_gC_g(-3-(4/3)C_g)+0.5\cos
k_x((8/3)u_g^2-(2/3)u^2-(2/3)uu_{2g}-(4/3)uu_d $

$+(8/3)uv_gC_g-(8/3)u_gv_gC_g+(2/3)uv_gC_g+(2/3)u_{2g}v_gC_g+(4/3)u_dv_gC_g$

$-(2/3)u_gvC_g-(2/3)u_gv_{2g}C_g-(4/3)u_gv_dC_g+4v_g^2C_g(-0.5-(2/3)C_g)$

$+4vv_gC_g-v^2C_g-vv_{2g}C_g-2vv_dC_g+v_gv(0.5C_g+(2/3)C_g^2)$

$+v_gv_{2g}C_g(0.5+(2/3)C_g)+2v_gv_dC_g(0.5+(2/3)C_g)$

$+(2/3)C_g(w_g^{(2)}+w_g^{(1)}+2w_g^{(3)}-4C_gw_g)+(4/3)W_{g3}^{(J)});$

\hangindent=1cm \noindent
$J_{55}=2C_g(v_g-v)+C_g\cos k_y(0.5v-2v_g+0.5v_{2g}+v_d);$

\hangindent=1cm \noindent
$%
J_{56}=(4/3)C_g(w_g-w)+(4/3)C_gu(v-v_g)+v_g^2(C_g+(4/3)C_g^2)+2v^2C_g+(4/3)W_{g4}^{(J)} 
$

$+vv_gC_g(-3-(4/3)C_g)+0.5\cos
k_y((8/3)u_g^2-(2/3)u^2-(2/3)uu_{2g}-(4/3)uu_d $

$+(8/3)uv_gC_g-(8/3)u_gv_gC_g+(2/3)uv_gC_g+(2/3)u_{2g}v_gC_g+(4/3)u_dv_gC_g$

$-(2/3)u_gvC_g-(2/3)u_gv_{2g}C_g-(4/3)u_gv_dC_g+4v_g^2C_g(-0.5-(2/3)C_g)$

$+4vv_gC_g-v^2C_g-vv_{2g}C_g-2vv_dC_g+v_gv(0.5C_g+(2/3)C_g^2)$

$+v_gv_{2g}C_g(0.5+(2/3)C_g)+2v_gv_dC_g(0.5+(2/3)C_g)$

$+(2/3)C_g(w_g^{(2)}+w_g^{(1)}+2w_g^{(3)}-4C_gw_g)+(4/3)W_{g3}^{(J)});$

\hangindent=1cm \noindent
$%
F0=-4uC_g-(8/3)uu_g+8C_guv+(8/3)wC_g-(8/3)w_gC_g+v_gu_g((32/3)C_g-4)+2w_g^{(1)} 
$

$%
+(8/3)v_guC_g+3v_gC_g-3vC_g-8v^2C_g-6v_g^2C_g-(8/3)v_g^2C_g^2+2vv_gC_g-8C_g^2vv_g 
$

$+(8/3)(W_1^{(\tau )}-W_{g1}^{(J)}-W_{g2}^{(J)});$

\hangindent=1cm \noindent
$F1=(2/3)u^2+u_gC_g-(10/3)uv_gC_g-2u_gvC_g+0.5u_gv+0.5uv_g+v^2C_g(0.75+C_g)$

$%
+v_g^2C_g(0.25+(5/3)C_g)+2vv_gC_g-0.5w_g-(2/3)w_g(C_{2g}+2C_d)+uvC_g-uv_gC_g 
$

$%
-(2/3)u_{2g}v_gC_g+(2/3)u_gvC_{2g}-(2/3)u_gvC_g+(2/3)uvC_{2g}+u_gv_gC_g+(2/3)uv_gC_{2g} 
$

$%
-u_gv_{2g}C_g-(2/3)uv_{2g}C_{2g}+(2/3)v^2C_g(C_g-C_{2g})+v_g^2C_g((5/3)C_g-(7/3)C_{2g}) 
$

$%
-(8/3)vv_gC_g^2+vv_{2g}C_g((-5/3)C_{2g}+C_g)+4v_gv_{2g}C_gC_{2g}+2(uvC_g-uv_gC_g 
$

$-(2/3)u_dv_gC_g+(2/3)u_gvC_d-(2/3)u_gvC_g+(2/3)uvC_d+u_gv_gC_g+(2/3)uv_gC_d$

$-u_gv_dC_g-(2/3)uv_dC_d+(2/3)v^2C_g(C_g-C_d)+v_g^2C_g((5/3)C_g-(7/3)C_d)$

$%
-(8/3)vv_gC_g^2+vv_dC_g((-5/3)C_d+C_g)+4v_gv_dC_gC_d)+2/3C_g(w_{g,2g}^{(1)}+2w_{gd}^{(1)}) 
$

$-(2/3)W_g^{(\tau )};$

\hangindent=1cm \noindent
$%
F2=3u_gv_gC_g+3uv_gC_g+(2/3)uv_gC_{2g}+(2/3)u_{2g}v_gC_g-(2/3)uv_{2g}C_{2g}-u_gv_{2g}C_g 
$

$-3uvC_g-(2/3)u_{2g}vC_g+(4/3)u_dv_gC_g-(4/3)uv_dC_d-2u_gv_dC_g-(4/3)u_dvC_g$

$+(4/3)uv_gC_d+0.5uv_g-0.5uv+C_g^2(v_{2g}+v+2v_d-4v_g)+v^2((-11/3)C_g^2$

$-0.25C_g+(2/3)C_{2g}C_g+(4/3)C_gC_d)+v_g^2((-50/3)Cg^2$

$-(7/3)(C_gC_{2g}-(14/3)C_gC_d-0.25C_g)+vv_g((52/3)C_g^2+0.5C_g$

$-(2/3)C_gC_{2g}-(4/3)C_gC_d)+v_gv_{2g}((8/3)C_g^2+(4/3)C_gC_{2g})$

$+vv_{2g}((-5/3)C_g^2+C_gC_{2g})+v_gv_d((8/3)C_gC_d+(16/3)C_g^2)$

$+vv_d((-10/3)C_g^2+2C_gC_d)+(1/6)W_g^{(J)};$

\hangindent=1cm \noindent
$J_{66}=F0+\gamma _g(F1+F2);$

\bigskip
\hangindent=1cm \noindent
$\widetilde{{\bf J}}^{(2)}{\bf -matrix}$

\smallskip
\hangindent=1cm \noindent
$Jd_{33}=-4C_d;$

\hangindent=1cm \noindent
$Jd_{34}=2C_d(v_d-v)+\cos k_xC_d(v_g+v_f-2v_g);$

\hangindent=1cm \noindent
$Jd_{35}=2C_d(v_d-v)+\cos k_yC_d(v_g+v_f-2v_g);$

\hangindent=1cm \noindent
$Jd_{36}=2C_d(v-v_d)+2u_d+(4/3)\gamma
_g(C_d(C_gv_g+C_fv_f)-2C_gC_dv_g+C_g(u_g+u_f)$

$-u_g(C_g+C_f));$

\hangindent=1cm \noindent
$Jd_{44}=2C_d(v_d-v)+\cos k_x(C_d(v_g+v_f)-2C_dv_g);$

\hangindent=1cm \noindent
$Jd_{55}=2C_d(v_d-v)+\cos k_y(C_d(v_g+v_f)-2C_dv_g);$

\hangindent=1cm \noindent
$%
Jd_{64}=(4/3)uC_d(v-v_d)+(4/3)C_d(w_d-w)+2v^2C_d+v_d^2(C_d+(4/3)C_d^2)+(4/3)W_{d4}^{(J)} 
$

$-vv_d(3C_d+(4/3)C_d^2)+\cos
k_x((4/3)C_d(uv_g-u_gv_d)+(4/3)u_gu_d-(4/3)C_dw_{2g}$

$%
+v_gv_d(C_d-(4/3)C_d^2)+2C_dv_g(v-v_d)-(2/3)u(u_g+u_f)+(2/3)C_d(v_d(u_g+u_f) 
$

$-u_d(v_g+v_f))+((2/3)C_d^2+0.5C_d)v_d(v_g+v_f)-C_dv(v_g+v_f)$

$+(2/3)C_d(w_{gd}^{(2)}+w_{gd}^{(1)})+(4/3)W_{d3}^{(J)});$

\hangindent=1cm \noindent
$%
Jd_{65}=(4/3)uC_d(v-v_d)+(4/3)C_d(w_d-w)+2v^2C_d+v_d^2(C_d+(4/3)C_d^2)+(4/3)W_{d4}^{(J)} 
$

$-vv_d(3C_d+(4/3)C_d^2)+\cos
k_y((4/3)C_d(uv_g-u_gv_d)+(4/3)u_gu_d-(4/3)C_dw_g$

$%
+v_gv_d(C_d-(4/3)C_d^2)+2C_dv_g(v-v_d)-(2/3)u(u_g+u_f)+(2/3)C_d(v_d(u_g+u_f) 
$

$-u_d(v_g+v_f))+((2/3)C_d^2+0.5C_d)v_d(v_g+v_f)-C_dv(v_g+v_f)$

$+(2/3)C_d(w_{gd}^{(2)}+w_{gd}^{(1)})+(4/3)W_{d3}^{(J)});$

\hangindent=1cm \noindent
$Fd0=-4uC_d-(8/3)uu_d+8C_duv+(8/3)wC_d-(8/3)w_dC_d+v_du_d((32/3)C_d-4)$

$%
+(8/3)v_duC_d+3v_dC_d-3vC_d-8v^2C_d-6v_d^2C_d-(8/3)v_d^2C_d^2+2vv_dC_d-8C_d^2vv_d 
$

$+2w_{dd}^{(1)}+(8/3)(W_1^{(\tau )}-W_{d1}^{(J)}-W_{d2}^{(J)});$

\hangindent=1cm \noindent
$%
Fd1=-(4/3)w_g(C_g+C_f)+(4/3)C_g(w_{gd}^{(1)}+w_{fd}^{(1)})+uv(4C_d+(4/3)(C_g+C_f)) 
$

$+(4/3)u_gv(C_g+C_f)-(8/3)u_dvC_g-4uv_dC_d+(4/3)uv_g(C_g+C_f)$

$-(4/3)u(v_gC_g+v_fC_f)-(4/3)v_dC_g(u_g+u_f)+(4/3)u_gv_gC_d+(8/3)u_dv_gC_g$

$%
-(4/3)u_dC_g(v_g+v_f)-(2/3)u_gC_d(v_g+v_f)+v^2((8/3)C_gC_d-(4/3)C_g(C_g+Cf)) 
$

$%
-(32/3)vv_gC_gC_d+2vC_gC_d(v_g+v_f)-(10/3)vC_d(v_gC_g+v_fC_f)+v_gv_d((20/3)C_gC_d 
$

$-2C_d(C_g+C_f))+(8/3)C_gv_g(v_gC_g+v_fC_f)+(16/3)v_dC_d(v_gC_g+v_fC_f)$

$-(8/3)v_g^2C_g(C_g+C_f)+vv_d((4/3)C_g(C_g+C_f)-(4/3)C_d(C_g+C_f));$

\hangindent=1cm \noindent
$Fd2=v^2(-(8/3)C_gC_d+(4/3)C_g(C_g+C_f))+16C_gC_dvv_g-4C_duv+2vv_gC_gC_d$

$+2vv_fC_dC_f-(10/3)vC_dC_g(v_g+v_f)-(4/3)C_g(u_g+u_f)v+(4/3)C_du_gv_g$

$+(8/3)C_gu_dv_g+(4/3)(C_g+C_f)uv_g-v_gv_d(20C_gC_d+2C_d(C_g+C_f))$

$-(8/3)C_g(C_g+C_f)v_g^2+vv_d((8/3)C_gC_d-(4/3)C_g(C_g+C_f))+4C_duv_d$

$%
+v_dC_d(2C_g(v_g+v_f)-2(v_gC_g+v_fC_f))+(4/3)C_g(u_g+u_f)v_d-(2/3)C_du_g(v_g+v_f) 
$

$%
-(4/3)u(C_gv_g+C_fv_f)-(4/3)C_gu_d(v_g+v_f)+(10/3)C_gC_dv_d(v_g+v_f)+(1/6)W_d^{(J)} 
$

$%
+2C_dv_d(C_gv_g+C_fv_f)+(8/3)C_gv_g(C_gv_g+C_fv_f)-4C_gC_dv_g+2C_gC_d(v_g+v_f); 
$

\hangindent=1cm \noindent
$Jd_{66}=Fd0+\gamma _g(Fd1+Fd2);$

\bigskip
\hangindent=1cm \noindent
$\widetilde{{\bf h}}{\bf -matrix}$

\smallskip
\hangindent=1cm \noindent
$h_{12}=2(0.5+\gamma _g+0.5\gamma _d);$

\hangindent=1cm \noindent
$h_{16}=2u(0.5+\gamma _g+0.5\gamma _d;$

\hangindent=1cm \noindent
$h_{26}=h_{16};$

\hangindent=1cm \noindent
$h_{33}=2(3/4+2C_g\gamma _g+C_d\gamma _d);$

\hangindent=1cm \noindent
$h_{34}=2(0.5u+u_g\gamma _g+0.5u_d\gamma _d);$

\hangindent=1cm \noindent
$h_{35}=h_{34};$

\hangindent=1cm \noindent
$h_{36}=2(-u+2C_g\gamma _g(v_g-v)+C_d\gamma _d(v_d-v));$

\hangindent=1cm \noindent
$h_{45}=h_{34};$

\hangindent=1cm \noindent
$h_{46}=2(-uv+\gamma _g(w_g-u_gv-v_gu)+0.5w+0.5\gamma _d(w_d-uv_d-u_dv));$

\hangindent=1cm \noindent
$h_{56}=h_{46};$

\hangindent=1cm \noindent
$h_{66}=2((3/4)u+2uv-w+\gamma
_g(2C_gu_g+2C_g(v^2+2v_gv)+(4/3)C_g^2(v^2+3v_g^2)$

$+2u^2-(2/3)u_g^2-8C_g((1/3)u_gv+uv_g))-(4/3)\gamma _gW_g^{(\tau )}+\gamma
_d(C_du_d+C_d(v^2+2vv_d)$

$+(2/3)C_d^2(v^2+3v_d^2)-4C_d((1/3)u_dv+uv_d)+u^2-(1/3)u_d^2)-(2/3)\gamma
_dW_d^{(\tau )});$

\bigskip
\hangindent=1cm \noindent
Above the following notations are taken

\medskip
\hangindent=1cm \noindent
$v=\frac 1N\sum\limits_{{\bf \kappa }}1;\quad \frac 1N\sum\limits_{{\bf %
\kappa }}\equiv \ \frac 1N\sum\limits_{{\bf \kappa \in \Omega }}$

\hangindent=1cm \noindent
$v_l=\frac 1N\sum\limits_{{\bf \kappa }}e^{i{\bf \kappa l}};\ {\bf %
l=g,d,2g,f;}$

\hangindent=1cm \noindent
$u=\frac 1N\sum\limits_{{\bf \kappa }}C_{{\bf \kappa }};$

\hangindent=1cm \noindent
$u_l=\frac 1N\sum\limits_{{\bf \kappa }}e^{i{\bf \kappa l}}C_{{\bf \kappa }%
};\ {\bf l=g,d,2g,f;}$

\hangindent=1cm \noindent
$w=\frac 1{N^{^2}}\sum\limits_{{\bf \kappa }_1{\bf ,\kappa }_2}C_{{\bf %
\kappa }_1-{\bf \kappa }_2};$

\hangindent=1cm \noindent
$w_l=\frac 1{N^{^2}}\sum\limits_{{\bf \kappa }_1{\bf ,\kappa }_2}e^{-i{\bf %
\kappa l}}C_{{\bf \kappa }_1-{\bf \kappa }_2};\ {\bf l=g,d,2g;}$

\hangindent=1cm \noindent
$w_{l_1l_2}^{(m)}=\frac 1{N^{^2}}\sum\limits_{{\bf \kappa }_1{\bf ,\kappa }%
_2}e^{-i{\bf \kappa }_1{\bf l}_1}e^{i{\bf \kappa }_2{\bf l}_2}C_{{\bf \kappa 
}_1-{\bf \kappa }_2};\ {\bf l}_{1,2}{\bf =g,d,2g,f;\ }$

$m=1:{\bf l}_1{\bf l}_2>0;\ m=2:{\bf l}_1{\bf l}_2<0;\ m=3:{\bf l}_1{\bf l}%
_2=0;$

\medskip
\hangindent=1cm \noindent
$w_l^{(m)}=w_{l_1l_2}^{(m)}(l_1=l_2);$

\medskip
\hangindent=1cm \noindent
$W_l^{(\tau )}=\frac 1{N^{^2}}\sum\limits_{{\bf \kappa }_1{\bf ,\kappa }_2%
{\bf ,\rho }}e^{-i({\bf \kappa }_1-{\bf \kappa }_2){\bf \rho }}e^{-i{\bf %
\kappa }_2{\bf l}}C_{{\bf \rho }}C_{{\bf \rho -l}};\ {\bf l=g,d,2g;}$

\hangindent=1cm \noindent
$W_1^{(\tau )}=\frac 1{N^{^2}}\sum\limits_{{\bf \kappa }_1{\bf ,\kappa }_2%
{\bf ,\rho }}e^{-i({\bf \kappa }_1-{\bf \kappa }_2){\bf \rho }}C_{{\bf \rho }%
}C_{{\bf \rho -g}};$

\hangindent=1cm \noindent
$W_2^{(\tau )}=\frac 1{N^{^2}}\sum\limits_{{\bf \kappa }_1{\bf ,\kappa }_2%
{\bf ,\rho }}e^{-i({\bf \kappa }_1+{\bf \kappa }_2){\bf g}}e^{-i({\bf \kappa 
}_1-{\bf \kappa }_2){\bf \rho }}C_{{\bf \rho }}C_{{\bf \rho -g}};$

\hangindent=1cm \noindent
$W_3^{(\tau )}=\frac 1{N^{^2}}\sum\limits_{{\bf \kappa }_1{\bf ,\kappa }_2%
{\bf ,\rho }}e^{-i{\bf \kappa }_1{\bf g}_x}e^{-i{\bf \kappa }_2{\bf g}_y}C_{%
{\bf \rho }}C_{{\bf \rho -g}};$

\hangindent=1cm \noindent
$W_{l1}^{(J)}=\frac 1{N^{^2}}\sum\limits_{{\bf \kappa }_1{\bf ,\kappa }%
_2}e^{i{\bf \kappa }_2{\bf l}}C_{{\bf \kappa }_1-{\bf \kappa }_2}C_{{\bf %
\kappa }_2};\ {\bf l=g,d};$

\hangindent=1cm \noindent
$W_{l2}^{(J)}=\frac 1{N^{^2}}\sum\limits_{{\bf \kappa }_1{\bf ,\kappa }%
_2}e^{i({\bf \kappa }_1-{\bf \kappa }_2){\bf l}}C_{{\bf \kappa }_1-{\bf %
\kappa }_2}C_{{\bf \kappa }_2};\ {\bf l=g,d};$

\hangindent=1cm \noindent
$W_{l3}^{(J)}=\frac 1{N^{^2}}\sum\limits_{{\bf \kappa }_1{\bf ,\kappa }%
_2}e^{-i{\bf \kappa }_2{\bf g}}(\gamma _l({\bf \kappa }_2)-\gamma _l({\bf %
\kappa }_1-{\bf \kappa }_2))C_{{\bf \kappa }_1-{\bf \kappa }_2}C_{{\bf %
\kappa }_2};\ \ {\bf l=g,d};$

\hangindent=1cm \noindent
$W_{l4}^{(J)}=\frac 1{N^{^2}}\sum\limits_{{\bf \kappa }_1{\bf ,\kappa }%
_2}(\gamma _l({\bf \kappa }_2)-\gamma _l({\bf \kappa }_1-{\bf \kappa }_2))C_{%
{\bf \kappa }_1-{\bf \kappa }_2}C_{{\bf \kappa }_2};\ {\bf l=g,d};$

\hangindent=1cm \noindent
$W_l^{(J)}=\frac 1{N^{^2}}\sum\limits_{{\bf \kappa }_1{\bf ,\kappa }%
_2}\sum\limits_{{\bf l}}(\sum\limits_{{\bf \rho }}e^{-i{\bf \kappa }_1{\bf %
\rho }}(C_{{\bf \rho +l}}C_{{\bf \rho -g}}-C_{{\bf \rho }}C_{{\bf \rho +l-g}%
})$

$-\sum\limits_{{\bf \rho }}e^{-i({\bf \kappa }_1-{\bf \kappa }_2){\bf \rho }%
}e^{-i{\bf \kappa }_1{\bf g}}C_{{\bf \rho }}C_{{\bf \rho +l+g}}+\sum\limits_{%
{\bf \rho }}e^{-i({\bf \kappa }_1-{\bf \kappa }_2){\bf \rho }}e^{-i{\bf %
\kappa }_1({\bf l}-{\bf g)}}C_{{\bf \rho }}C_{{\bf \rho +l-g}});\ {\bf l=g,d.%
}$


\end{document}